\documentclass[aps,pra,floats,floatfix,twocolumn]{revtex4}

\usepackage{ifthen}
\usepackage{ifpdf}
\usepackage{color}

\ifpdf
\usepackage{graphicx}
\usepackage{epstopdf}
\else
\usepackage{graphicx}
\usepackage{epsfig}
\fi
\graphicspath{{./Figs/}{./}}

\usepackage{latexsym}
\usepackage{amsmath}
\usepackage{amssymb}
\usepackage{bm}
\usepackage{wasysym}

\newcommand{\BraKet}[3]{\left\langle #1 \middle| #2 \middle| #3 \right\rangle}

\newcommand{\beq}{\begin{eqnarray}}
\newcommand{\eeq}{\end{eqnarray}} 

\newcommand{\hide}[1]{}
\newcommand{\Eq}[1]{\textcolor{blue}{{Eq.}\!\!~(\ref{#1})}}

\newcommand{\Fig}[1]{\textcolor{blue}{Fig.}\!\!~\ref{#1}}
\definecolor{myred}{rgb}  {0.5,0.0,0.0}
\newcommand{\rmrk}[1]{\textcolor{black}{#1}}

\newcommand{\sect}[1]{{\bf #1.--}}

\begin{document}
\title{Many-body adiabatic passage: Quantum detours around chaos}
\author{Amit Dey$^1$}
\author{Doron Cohen$^2$}
\author{Amichay Vardi$^1$} 
\affiliation{$^1$ Department of Chemistry,
Ben-Gurion University of the Negev, Beer-Sheva 84105, Israel}
\affiliation{$^2$ Department of Physics,
Ben-Gurion University of the Negev, Beer-Sheva 84105, Israel}
\date{\today}

\begin{abstract}
We study the many-body dynamics of stimulated Raman adiabatic passage in the presence of on-site interactions. In the classical mean-field limit, explored in Phys. Rev. Lett. {\bf 121}, 250405 (2018), interaction-induced chaos leads to the breakdown of adiabaticity under the quasi-static variation of the parameters, thus producing {\em low} sweep rate boundaries on efficient population transfer. We show that for the corresponding many-body system,  alternative quantum pathways from the initial to the target state,  open up at even slower sweep rates. These quantum detours avoid the chaotic classical path and hence allow a robust and efficient population transfer.
\end{abstract}
\maketitle

\section{Introduction}

Adiabatic passage is a major tool of quantum control and quantum state engineering. For two-level systems, the Landau-Zener-Stuckelberg-Majorana (LZSM) linear crossing  \cite{Landau32,Zener32, Stuckelberg32,Majorana32} has been a prominent paradigm. Three level configurations offer, in addition to Landau-Zener-like 'rapid adiabatic passage'  (RAP) schemes, also the possibility of stimulated Raman adiabatic passage (STIRAP) \cite{Gaubatz90, Vitanov17} in which an interference-induced dark state allows for the efficient transfer of population from source to target state without projection onto an intermediate (often spontaneously decaying) state.

\subsection{Nonlinear Adiabatic Passage}
Advances in the field of Bose-Einstein condensation (BEC) of dilute atomic gases, have triggered great interest in generalizations of adiabatic passage scenarios to many-body interacting systems. Interactions were shown to have a dramatic effect on two-mode ('Bose-Hubbard dimer') sweep physics,  with power-law dependence of remanent populations on sweep duration and finite nonadiabatic fractions at slow driving rates beyond a critical interaction strength. Interaction-induced effects were obtained both in the classical nonlinear mean-field limit \cite{Zobay00,Wu00,Liu02,Liu03,Witthaut06,Witthaut11,Ishkhanyan10} and in semiclassical or quantum many-body treatments \cite{Witthaut06,Trimborn10,Altland08,Mannschott09,Chen11,Paredes13}. Similar behavior was obtained for coupled atomic and molecular condensates \cite{Javanainen99,Yurovsky00, Heinzen00, Ishkanyan04, Altman05, Pazy05, Tikhonenkov06, liu08, liu08b}. A common denominator for all these studies is the connection between the nonlinear effect and energetic stability. The unique nonlinear behavior is universally attributed to the emergence of a separatrix, containing an hyperbolic instability.

The effects of interactions on three-mode adiabatic schemes \cite{Graefe06,Rab08,Bradly12,Polo16,Dupont-Nivet15} are not as well studied as their two-mode counterparts. An attempt to extend the two-mode idea that nonadiabaticity can be determined  from energetic stability analysis, has been made in Ref.~\cite{Graefe06}. Mean field adiabatic stationary points (SPs) for nonlinear RAP and STIRAP were found and the appearance of a so called 'horn' avoided crossing in the adiabatic energy diagram was consequently linked to the loss of adiabaticity at low sweep rates.  However, as pointed out in \cite{Dey18},  the three-mode system offers richer physics than its two-mode counterpart, due to its inherent nonintegrability. Specifically, energetic stability of a SP is insufficient to determine its {\em dynamical} stability because classical trajectories can diverge within a single multidimensional energy surface, containing both chaotic and quasi-integrable domains \cite{Arwas15}. Consequently, adiabatic passage efficiency may be affected by the appearance of phasespace structures which are not manifested in the stationary point energy diagrams. The analysis of adiabatic passage involving such structures goes well beyond the LZSM paradigm or any of its nonlinear extensions. 

\subsection{Nonlinear STIRAP through chaos}
In Ref.~\cite{Dey18}, we have shown that the loss of adiabaticity during nonlinear STIRAP with repulsive interaction, is not related to energetic instability. Instead, adiabaticity breaks down at slow sweep rates due to dynamical instability which has no trace in the energy bifurcation diagram. This dynamical instability corresponds to the embedding of the followed SP in chaotic strips that emerge within its energy surface. The outcome of this novel breakdown mechanism, is that classical adiabaticity may be restored by {\em faster} variation of the control parameter, so as to guarantee that the chaotic interval is traversed before ergodization takes place. Thus, in addition to the standard upper sweep rate boundary required to ensure efficient adiabatic transfer, there exists also a {\em lower} sweep rate boundary that ensures successful passage through chaos. This lower boundary increases with the interaction until it coincides with the higher boundary, thus making efficient population transfer impossible beyond a critical interaction strength.

\subsection{Quantum detours around chaos}
In this work, we go beyond the classical mean-field picture and study the many-body quantum dynamics of STIRAP in the presence of inter-particle interactions. Similarly to the quantum two-mode case, classical adiabaticity corresponds to a series of diabatic transitions through avoided crossings between the many-body adiabatic eigenstates. This {\em classically adiabatic} path is interrupted during chaotic intervals when the followed SP becomes dynamically unstable and the corresponding many-body state loses one-particle coherence. However, due to a diabatic-to-adiabatic transition through a single many-body avoided crossing, new quantum detours open up, which avoid the classically chaotic regions in phase space and reenable efficient population transfer from the initial to the target mode.

\sect{Outline}
The quantum many-body model is presented in Section~II.   Numerical many-body simulations are presented and interpreted in terms of 'adiabatic passage through chaos' and 'quantum detours' in Section~III. Conclusions are provided in Section~IV.


\begin{figure}
\centering
\includegraphics[width=\hsize]{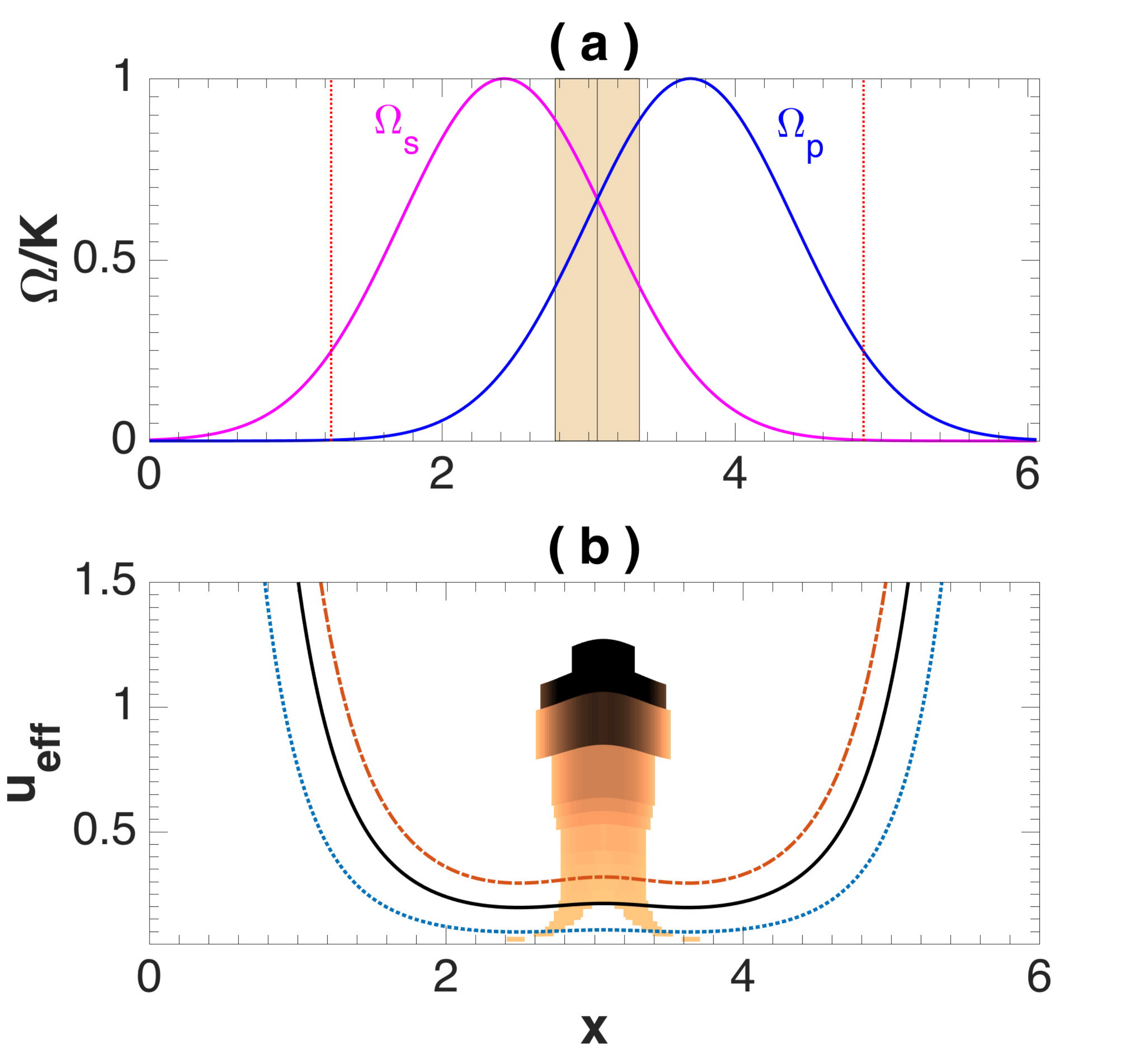} 
\caption{(color online) 
Many-body STIRAP: 
{\bf (a)} The STIRAP pulse scheme with the Stokes pulse preceding the pump.  
Here and below, the classically chaotic intervals 
and avoided horn crossings are marked by shaded regions and vertical dotted lines, respectively. 
Here they are shown for $u=0.2$. The detuning here and throughout the manuscript is $\varepsilon=0.1$.
{\bf (b)} The effective nonlinearity $u_{\text{eff}}(x)$ 
for ${u=0.1}$ (dotted blue), and $0.2$ (solid black), and $0.3$ (dash-dotted red). 
The marked region corresponds to the magnitude of the imaginary part 
of the characteristic Bogoliubov frequencies around the followed classical SP, 
i.e. to the region of dynamical instability: The followed SP is embedded 
in chaos when $u_{\text{eff}}$ is inside the shaded region.}
\label{scheme}
\end{figure}

\section{The Many-Body Model}
\label{model}

\subsection{The Bose-Hubbard Trimer Hamiltonian}
The many-body dynamics of STIRAP in the presence of on-site interaction, is modeled by the time-dependent Bose-Hubbard trimer Hamiltonian \cite{Graefe06,Eilbeck95,Hennig95,Franzosi03,Flach97,Nemoto00,Franzosi02,Hiller06,Tikhonenkov13} for $N$ particles in three second-quantized modes, that can
be either spatial lattice sites or internal atomic states :
\begin{eqnarray}
\label{ham}
\mathcal{H}&=& \ {\cal E}\hat{n}_2 \ + \ \frac{U}{2}\sum_{j=1}^3\hat{n}^2_i  \nonumber \\
~&-& \frac{1}{2}\left(\Omega_p(x)\hat{a}_2^{\dagger}\hat{a}_1 + \Omega_s(x)\hat{a}_3^{\dagger}\hat{a}_2+h.c.\right)~.
\end{eqnarray}
Here,  $\hat{a}_j$, $\hat{a}_j^\dag$ are bosonic operators with associated occupations $\hat{n}_j\equiv \hat{a}_j^{\dagger}\hat{a}_j$.  The interaction strength is~$U$, while ${\cal E}$ is the middle site bias, equivalent to the one-photon detuning of the optical scheme \cite{Gaubatz90, Vitanov17}. The couplings are Gaussian Stokes and Pump pulses 
\beq
\Omega_{s,p}(x) \ \ = \ \ Ke^{-\left(x-x_{s,p}\right)^2}
\eeq
which depend on the dimensionless parameter~$x$. The standard realization is a simple constant-rate sweep $x(t) = t/\tau$, with a `counterintuitive' sequence ${ x_p-x_s > 0 }$, as shown in \Fig{scheme}a. 

The dimensionless characteristic parameters of the the trimer Hamiltonian of \Eq{ham} are the interaction $u=UN/K$, the couplings $\kappa_{p,s}=\Omega_{p,s}/K$, and the detuning $\varepsilon={\cal E}/K$. We also define an effective interaction parameter,
\begin{equation}
u_{\text{eff}}(x) \ = \ \frac{UN}{\sqrt{\Omega_p^2(x) + \Omega_s^2(x)}}
\end{equation}
that reflects the~$x$ dependence of the couplings. 
This quantity is largest at the early and late stages of the process, 
when the linear coupling terms are small, see \Fig{scheme}b.

\begin{figure}
\centering
\includegraphics[width=\hsize]{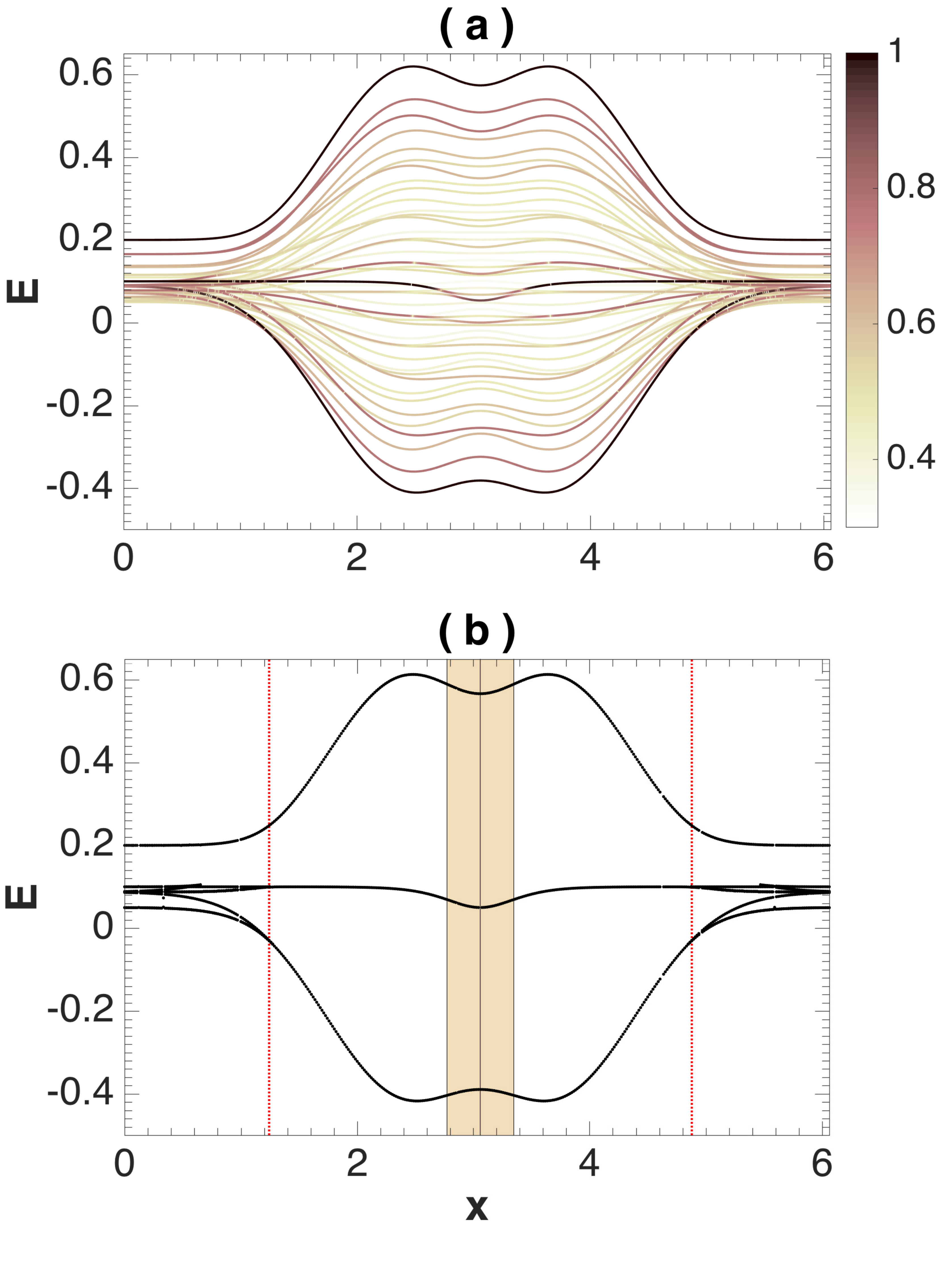} 
\caption{(color online) 
Quantum vs. classical spectra.  
{\bf (a)} The many-body adiabatic eigenenergies $E_{\nu}(x)$ 
for an ${N=8}$ particle system with $u=2\varepsilon=0.2$. 
Each line is color-coded according 
to the one-particle purity $\gamma$ of the eigenstate. 
{\bf (b)} The energies $E_{\text{SP}}(x)$ of the classical SPs. 
In the absence of chaos the SPs can support coherent many-body eigenstates. \\ }
\label{spectra}
\end{figure}
%
%


\subsection{Classical Stability and Chaos}
In the classical limit of the trimer Hamiltonian (see Appendix~A) the field operators $\hat{a}_j$ are replaced by $c$-numbers $a_j=\sqrt{n_j}e^{i\phi_j}$. The classical motion thus has three degrees of freedom, with $\{n_j,\phi_j\}_{j=1,2,3}$ serving as conjugate action-angle coordinates. Due to conservation of $N$, the clasical  phase-space reduces to two freedoms (two population imbalances and two conjugate relative phases). Adiabatic classical motion  corresponds to preparing the system near one of the stationary points in this 4D phasespace and following it, 
as it translates due to the slow variation of $x$,  to the target state. Specifically, in linear STIRAP the followed SP is, 
\begin{eqnarray}
n_1-n_3&=&N\left[\cos^2\vartheta(x)-\sin^2\vartheta(x)\right],\nonumber\\
\phi_1-\phi_3&=&\pi,\nonumber\\
n_2&=&0~,
\end{eqnarray} 
translating from the first mode to the third mode as the mixing angle $\vartheta(x)\equiv\tan^{-1}(\Omega_p(x)/\Omega_s(x))$ varies from $0$ to $\pi/2$. 

Similar adiabatic paths from the initial state to the target state, can be found in the presence of moderate interaction \cite{Graefe06,Dey18}. The nonlinearity of the classical equations of motion with any given $u\neq0$, means we should study the {\em stability} of the followed SP  at any value of~$x$ during its evolution. 
This can be done using either Bogoliubov analysis (Appendix~B) or via numerical simulation (Appendix~C). 
We distinguish between {\em energetic} stability and {\em dynamical} stability.
The former term applies if the SP is situated at a local minimum or a local maximum 
of the energy landscape. In \Fig{scheme}a, energetic stability is lost, for the pertinent parameters, in between
the two vertical dotted lines. These lines correspond to the avoided {\em horn crossing} of Ref.~\cite{Graefe06}.    

By contrast, dynamical instability is indicated by complex Bogoliubov frequencies. In our case, 
dynamical instability is associated with the embedding of the followed SP in chaotic strips in phase space (Appendix~C). 
The imaginary part of the Bogoliubov frequencies thus corresponds to a rate ${1/t_s}$. where $t_s$ is a characteristic 
time for spreading within the chaotic strip.   The marked region in \Fig{scheme}b images this rate at various interaction strengths.
For any given $u$, the chaotic interval during which the SP is embedded in chaos, is determined from the $x$ range where
$u_{\mathrm{eff}}$ is inside this dynamical instability region. For example, the shaded interval in \Fig{scheme}a corresponds to the 
duration where the $u=0.2$ solid line of \Fig{scheme}b enters the unstable region. For any given $u$ we denote the width of this chaotic interval as $\xi_s$.

\begin{figure}
\centering
\includegraphics[width=\hsize]{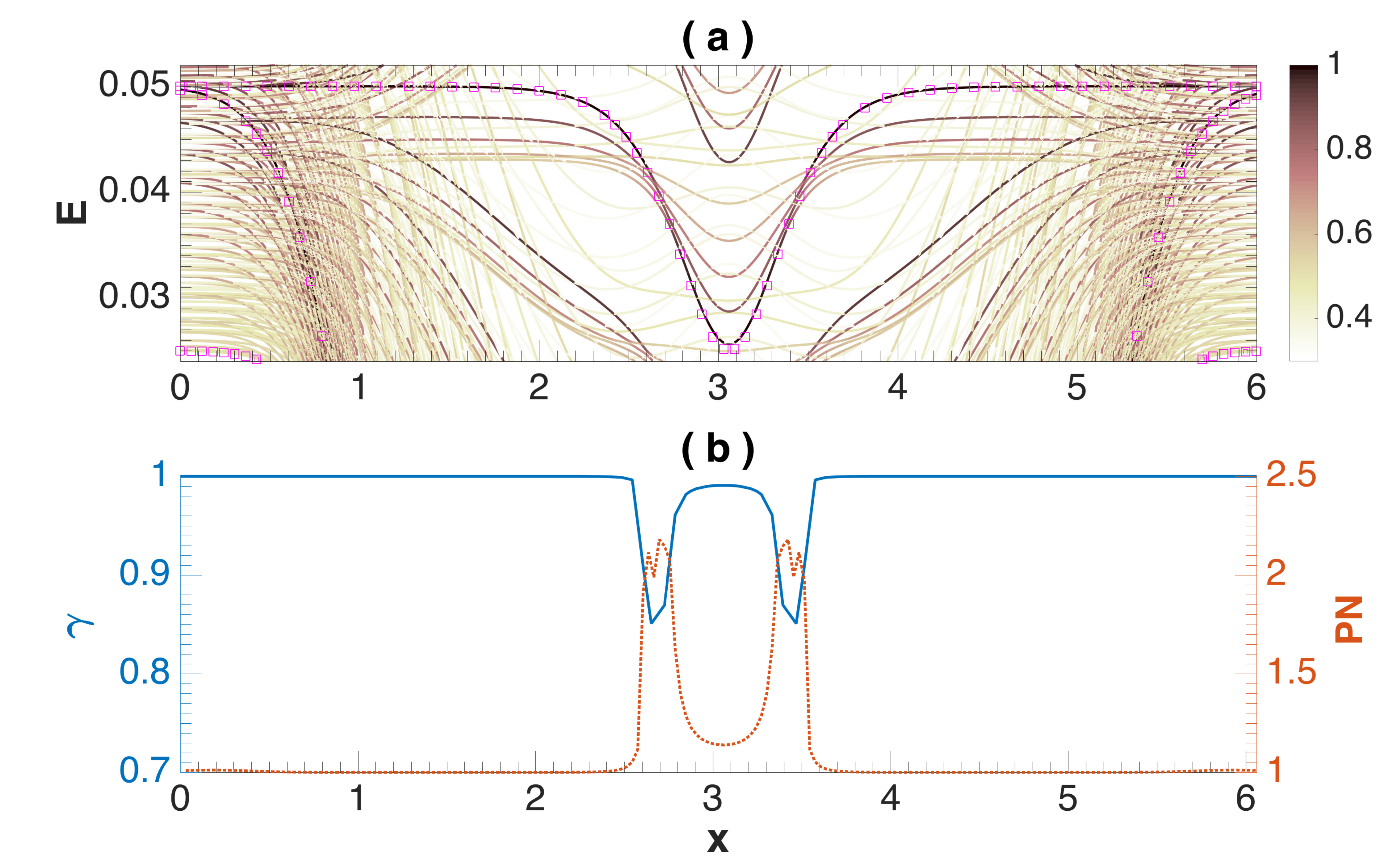} 
\caption{(color online) 
{\bf (a)} The many-body adiabatic eigenenergies $E_{\nu}(x)$ 
for \rmrk{${N=30}$}  and $u=0.1$. Lines are color-coded as in \Fig{spectra}. Magenta square markers indicate the energy of the followed classical SP.  {\bf (b)} The one-particle purity of the many body 
adiabatic eigenstates along the classical path (upper blue line) and the participation number of coherent states corresponding to the followed classical SP (lower red line). }
\label{coherence_rep}
\end{figure}

\begin{figure*}
\centering
\includegraphics[width=\hsize]{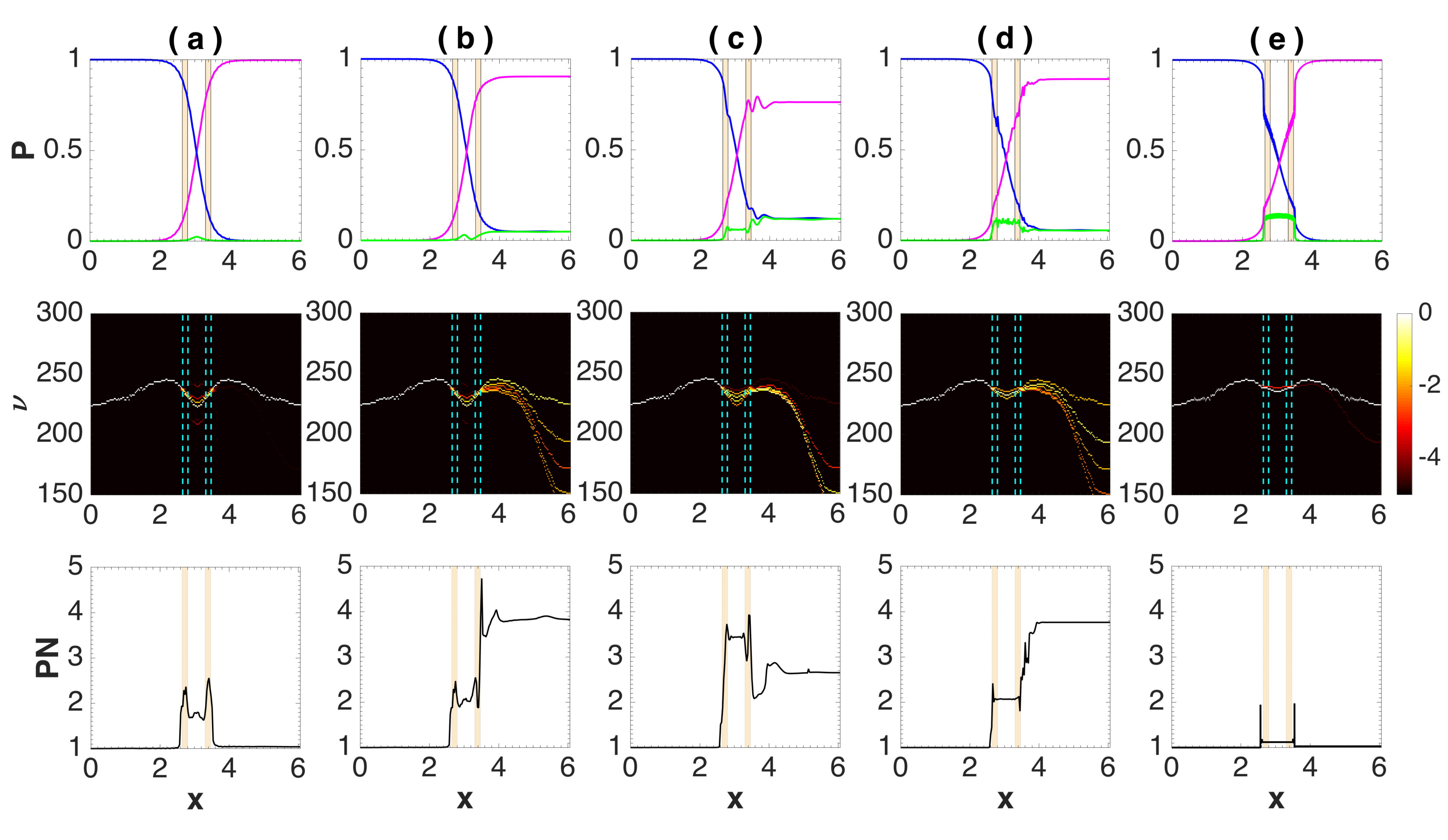} 
\caption{(color online)  
Site population dynamics $P_i(t)$ (top), many-body population distribution $p_\nu(t)$, 
and the corresponding participation number PN$(x)$, 
for $N=30$ system with ${u=0.1}$, and ${\varepsilon=0.1}$.
The plots are against the time dependent parameter~$x(t)$. 
The sweep rate is: 
(a) ${\dot x}/K=8.6\times 10^{-3}$; 
(b) $6\times 10^{-3}$;
(c) $1.2\times 10^{-3}$;  
(d) $2\times10^{-4}$;
and (e) $6\times10^{-6}$. 
Shaded regions and vertical lines mark the chaotic intervals.}
\label{dynamics}
\end{figure*}

\subsection{Eigenstates and their purity}
Due to total number conservation, the Hilbert space of the quantum many-body system is spanned by the Fock basis states,
\begin{equation} \nonumber
\left | n, n_2\right\rangle=\frac{1}{\sqrt{N !}}  \left(\hat{a}_1^\dag\right)^{\frac{N{-}n_2{+}n}{2}}  \left(\hat{a}_2^\dag\right)^{n_2}   \left(\hat{a}_3^\dag\right)^{\frac{N{-}n_2{-}n}{2}}  \left| \mathrm{vac} \right\rangle~,
\end{equation}
where $n_2=0,1,...,N$ is the intermediate site's occupation and $n=n_1-n_3=-(N-n_2),...,N-n_2$  is the population imbalance between the outer sites. Representing the Hamiltonian in this basis, we find the many-body adiabatic eigenstates $|\nu\rangle$ that diagonalize the Hamiltonian at each instant $x$,
\begin{equation}
\mathcal{H}(x) |\nu(x)\rangle \ = \ E_\nu(x) |\nu(x) \rangle~.
\end{equation}
The one-particle coherence of any many-body state $|\Psi\rangle$, is quantified by its purity 
\begin{equation}
\gamma=\mathrm{Tr}\left([\rho^{(sp)}]^2\right)~,
\end{equation} 
where $\rho^{(sp)}_{i,j} \equiv (1/N) \langle \Psi | \hat{a}_i^\dag \hat{a}_j |\Psi\rangle$ 
is the reduced single-particle density matrix. The trimer's coherent states are characterized by $\gamma=1$ whereas completely incoherent states have $\gamma=1/3$.  In \Fig{spectra}a, we plot the many-body adiabatic energies as a function of $x$, color-coded according to the one-particle coherence of the corresponding eigenstates.

\subsection{The participation number}
Given an arbitrary many-body state $|\Psi\rangle$, 
we can expand it in the many-body adiabatic eigenstate and define 
\beq
p_{\nu} \ \ = \ \ \left|\langle\nu|\Psi\rangle\right|^2
\eeq
This is essentially the local density of states (LDOS) with respect to the reference~$\Psi$. 
From this distribution one can extract the {\em participation number},
\begin{equation}
\mathrm{PN}=\frac{1}{\sum_\nu p_\nu^2}~.
\label{participation}
\end{equation}
This quantity corresponds to the number of adiabatic eigenstates that participate 
in the wavepacket $|\Psi\rangle$. 

In a strictly adiabatic STIRAP process the time dependent state $\Psi(t)$ 
is an instantaneous eigenstate of the $\mathcal{H}(x)$ Hamiltonian 
at any moment. The time dependent PN is calculated in the adiabatic basis. 
Strict adiabaticity means that  ${\mathrm{PN}(t) \approx 1}$ at any moment.

\section{Quantum dynamics}
\subsection{Numerical simulations}
Consider first the case of linear STIRAP ($u{=}0$).  
The system is prepared with all particles occupying the first mode, 
namely $|\Psi(x=0)\rangle=\left|N,0\right\rangle$. 
The adiabatic sweep from $x=0$ to $x=x_f$ transfers the population 
to the third mode ($|\Psi(x=x_f)\rangle\approx\left|-N,0\right\rangle$) by following 
the coherent dark eigenstate,
\begin{equation}
\label{ds}
\left| \text{SP} \right\rangle_x = \frac{1}{\sqrt{N !}}  \left(\cos{\vartheta(x)} \hat{a}_1^\dag - \sin{\vartheta(x)} \hat{a}_3^\dag \right)^N \left| \mathrm{vac} \right\rangle~.
\end{equation}
The dark state (\ref{ds}) does not project on the intermediate mode 
at any time ($\langle n_2 \rangle_x=0$). This state is labeled "SP", 
because it corresponds to a semiclassical minimal wavepacket, 
supported by the classical SP whose energy 
is ${E[\text{SP}]= 0 }$.

\begin{figure}[t]
\centering
\includegraphics[width=\hsize]{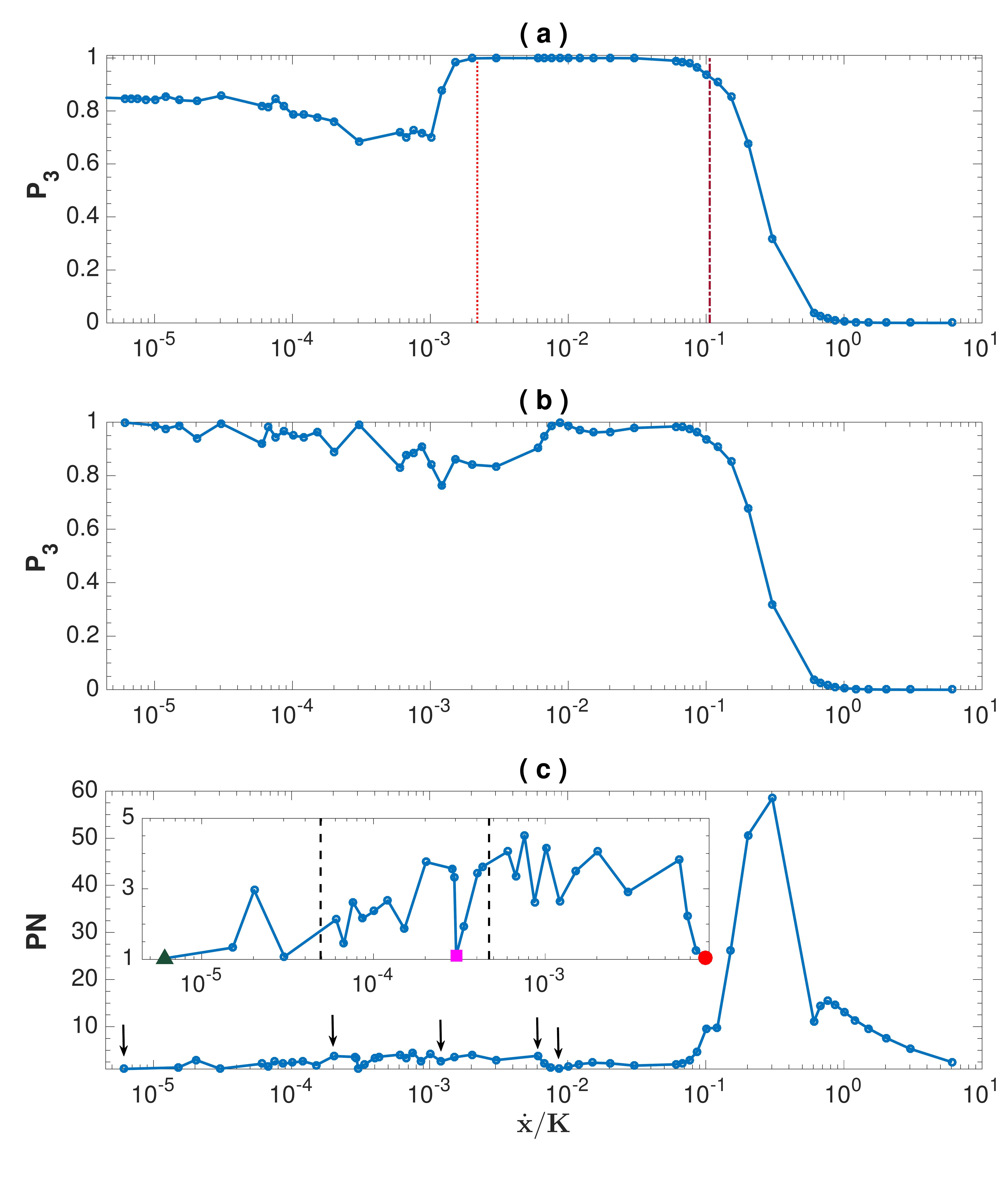}
\caption{(color online)  
The dependence of STIRAP efficiency on the rate of the sweep: 
(a) classical, with the predicted boundaries of \Eq{apc} marked by verical dotted and dash-dotted lines; (b) quantum; (c) PN of the final quantum $p_\nu$ distribution. Arrows mark the location of the simulations shown in \Fig{dynamics}a-e. The vertical dashed lines in the inset mark the expected detour thresholds of \Eq{mba}. The red circle marks the classical adiabatic dynamics of \Fig{pathways}a, whereas the magenta square and green triangle mark respectively, the quantum detours of \Fig{pathways}b and \Fig{pathways}c.}
\label{summary}
\end{figure}

As shown in \Fig{spectra}b,  a similar SP (the classical state whose energy is initially $E[\text{SP}]=u/2$) is followed in the case of nonlinear STIRAP with repulsive interaction ($u>0$). However, unlike in the linear case, one particle coherence is not maintained when this point becomes dynamically unstable. \Fig{coherence_rep} shows a close-up on the many-body adiabatic spectrum in the vicinity of the followed classical SP's energy. The one-particle purity along the classical adiabatic path is close to unity, except during the classically chaotic intervals. This is so because when the dynamics becomes chaotic,  the phase-space distribution of the many-body eigenstate is spread nearly uniformly throughout  the chaotic strip, and does not resemble a localized coherent state. Similarly, if we plot the participation number of a coherent state $\left| \text{SP} \right\rangle_x$  that corresponds to the followed classical SP (slightly different from \Eq{ds} due to the nonlinearity), we see that it increases during the chaotic intervals, because this localized state projects equally on all the $|\nu\rangle$ eigenstates that are supported by the chaotic strip. Thus, similarly to the classical case, we expect that in order to avoid spreading over these states, the chaotic intervals would have to be traversed fast with respect to the spreading time $t_s$.

In \Fig{dynamics}, we plot the results of numerical propagation of the many-body system with the Hamiltonian \Eq{ham} at different sweep rates. In the first row we plot the mode populations:
\beq
P_i(t) \ \ = \ \ \frac{1}{N}\BraKet{\Psi(t)}{a_j^{\dag}a_j}{\Psi(t)} ~,
\eeq  
while in the second and third rows we plot the instantaneous many-body population distribution $p_{\nu}$, and the corresponding participation number $\mathrm{PN}$. All quantities are plotted against~$x(t)$. 
Looking at columns \Fig{dynamics}a-\Fig{dynamics}c we see that slower sweep results in {\em enhanced} spreading of population between many-body eigenstates during the chaotic intervals, accompanied by the corresponding increase of the participation number. However, as shown in \Fig{dynamics}d-\Fig{dynamics}e, further slowing {\em recovers} efficient transfer and unit participation. This recovery is absent in the classical simulations and is thus a pure quantum many-body effect.

The dependence of STIRAP efficiency $P_3(\infty)$ and the corresponding PN$(\infty)$ on the sweep rate $\dot{x}$, is summarized in \Fig{summary}. We now turn to provide detailed explanation for the observed $\dot{x}$ dependence. The fastest sweep r.h.s of \Fig{summary} corresponds to the sudden limit, where no population is transferred and $\mathrm{PN}\sim 1$ simply because the system remains in the initial state. The large participation number hump is the standard sudden~$\rightarrow$~adiabatic transition, with unit participation attained again when there is classical adiabatic passage (see \Fig{dynamics}a).  However, interesting new features emerge at lower sweep rates, where the transfer efficiency {\em decreases} and the final participation {\em increases} as the sweep becomes slower. This enhanced spreading at moderately slower sweep rates is  equivalent to the semiclassical  'passage through chaos' analysis of Ref.~\cite{Dey18}. We review the main results in  Section~III.B.  By contrast, the quantum recovery at the slowest presented sweep rates (l.h.s of \Fig{summary}b and \Fig{summary}c, see also \Fig{dynamics}e) has no classical equivalent (compare the l.h.s. of the quantum \Fig{summary}b and the classical \Fig{summary}a). Its explanation by the emergence of 'quantum detours' around the classically chaotic regions, is provided in Section~III.C.

\subsection{Passage through chaos}

\begin{figure}[t]
\centering
\includegraphics[width=\hsize]{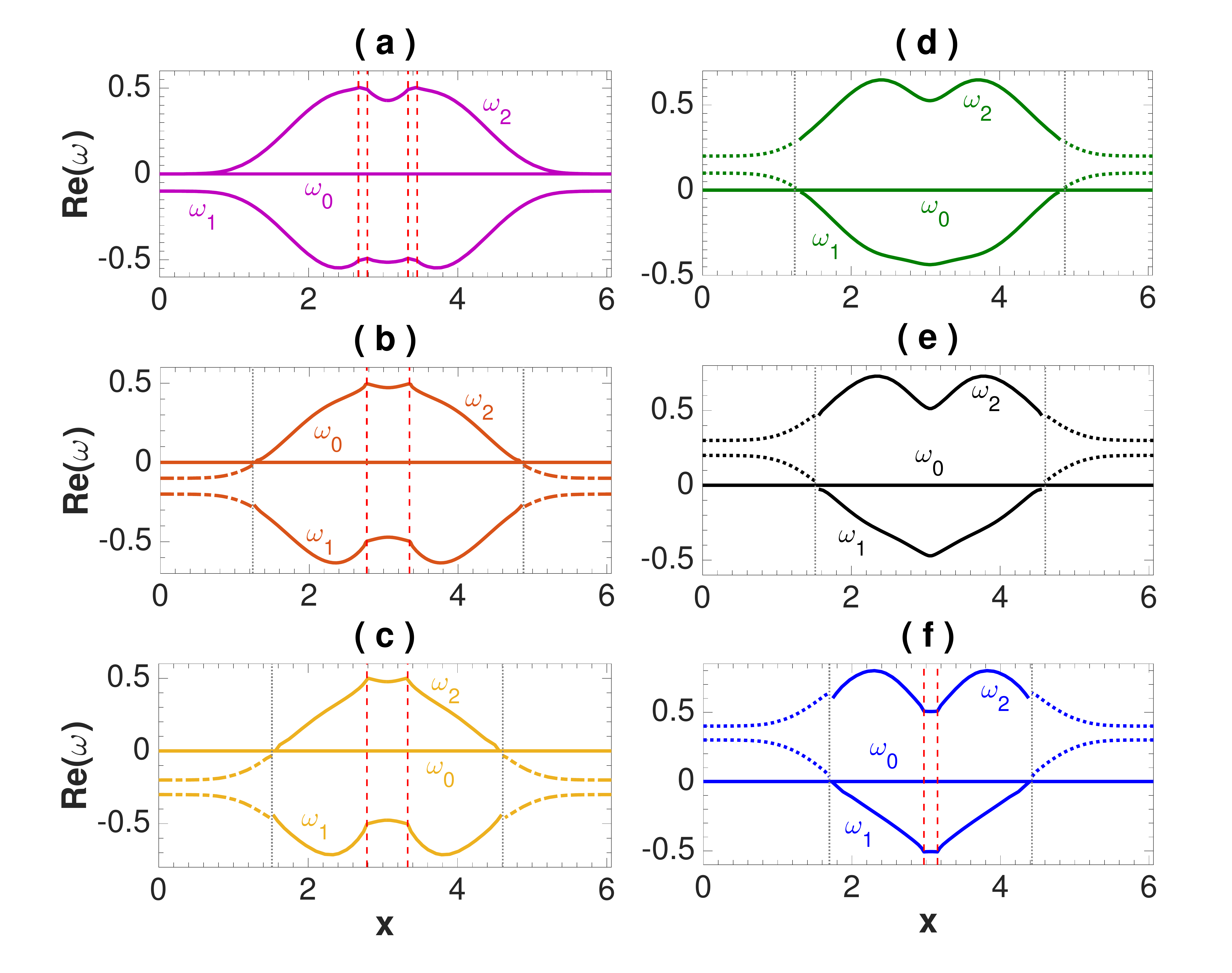} 
\caption{(color online) 
Real part of the Bogoliubov frequencies for:
(a) $u=0.1$;
(b) $0.2$; 
(c) $0.3$;
(d) $-0.1$; 
(e) $-0.2$;
and (f) $-0.3$. 
Dotted vertical lines mark the horn crossings (if exist), 
where one of the frequencies changes sign, indicating 
that energetic stability is lost. 
Dashed vertical lines indicate the borders 
of the dynamically unstable interval,  
where the frequencies are complex. \\ }
\label{bog_r}
\end{figure}

\begin{figure}[t]
\centering
\includegraphics[width=\hsize]{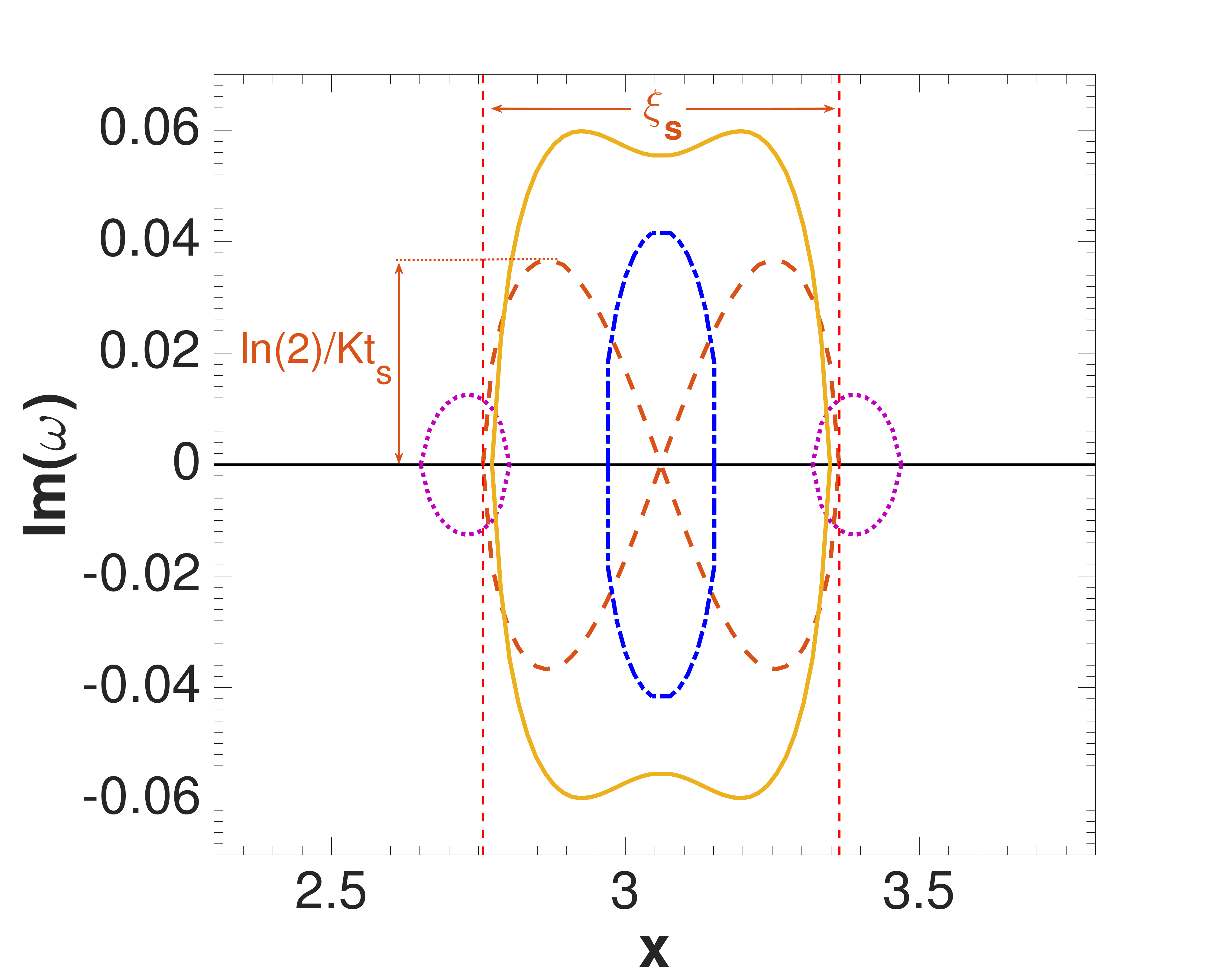} 
\caption{(color online) 
Imaginary part of the Bogoliubov frequencies for $u=0.1$ (dotted magenta); $0.2$ (dashed red); $0.3$ (solid orange); $-0.3$ (dash-dotted blue). 
The imaginary part of $\omega_{1,2}$ when $u=-0.1$ and $u=-0.2$ is identically zero (not plotted). 
The frequencies are complex within an interval of width~$\xi_s$. 
The instability time $t_s$ is determined by the maximal value of $\mathrm{Im}(\omega)$. The extraction of these two parameters is illustrated 
for the $u=0.2$ curve.}
\label{bog_i}
\end{figure}

The Bogoliubov transformation (Appendix~B) approximates the Hamiltonian in the vicinity of the SP by a quadratic form, namely,   
\begin{equation} \label{hamq} 
{\cal H} \ \ = \ \ \sum_{j=1}^d \omega_j \hat{c}_j^\dag \hat{c}_j~,
\end{equation}
where $d$ is the number of degrees of freedom. For the two-freedoms Bose-Hubbard trimer, there are two non-vanishing frequencies $\omega_{1,2}$. Their real part is plotted in \Fig{bog_r}  for all $x$ throughout the STIRAP evolution, at different values of the interaction parameter $u$. 
The imaginary part of $\omega_{1,2} $ for the same parameters, is plotted in \Fig{bog_i}. 
The chaotic interval, within which the SP is dynamically unstable, 
is identified as the range where the latter is non-zero.

%

{\em Energetic Stability} is more strict than {\em dynamical stability}.
It is determined by the signs of $\omega_{1,2}$ when they are real.
If both frequencies are positive (negative), the SP is an energy minimum (maximum). Opposite signs indicate a saddle-point in the energy landscape, but note again that
this  does not imply that the followed SP is an hyperbolic point in phase space, because the motion {\em within} a single multidimensional (3D in our case) energy surface can still be elliptic ($\mathrm{Im} (\omega_j)=0$) or hyperbolic ($\mathrm{Im} (\omega_j)\neq 0$).

\rmrk{For repulsive interaction, the followed SP  at early and late times 
where the Hamiltonian is interaction dominated, is a self-trapped maximum when $u>\varepsilon$ (interaction energy is lost by transfer of particles from the initial site to either of the unpopulated sites) or an energy saddle point if $0<u<\varepsilon$ (energy is gained by moving particles to the detuned central site and lost by transfer to the target site). For $0<u<\varepsilon$ (panel~\Fig{bog_r}a) the SP remains a saddle} at all times. For $u>\varepsilon$ (panels~\Fig{bog_r}b,c)
we have horn crossing \cite{Graefe06}, meaning that a local maximum becomes a saddle-point.   
By contrast, for attractive interaction, the SP starts out as a minimum and the horn crossing, 
manifested as a minimum to saddle-point transition, exists for any ${u<0}$ (panels~\Fig{bog_r}d-f).

The dynamical instability appears for $u>\varepsilon/2$ \cite{Dey18}, due to the  embedding of the followed SP in a chaotic strip (Appendix C).    
The implication is a lower threshold for the sweep rate required to maintain 
adiabaticity. If the sweep is too slow, the wavepacket has the time to spread over 
the chaotic strip, coherence is lost, and the terminal transfer efficiency is spoiled.      
We define a characteristic spreading time as follows:
\begin{equation}
t_s = \frac{\ln{2}}{\max\left[ \mathrm{Im}(\omega)\right]}~,
\end{equation}
It is the time that it takes for the dispersion around the SP to be doubled. 
The chaotic interval's width $\xi_s$, is determined by the condition $\mathrm{Im}(\omega)\neq 0$, 
see \Fig{bog_i}. The draining of the SP region can be avoided if $\xi_s$ is traversed 
on a shorter time scale than $t_s$ \rmrk{(see Appendix~D)}. 
%
%
Combining the lower sweep rate threshold with the standard upper sweep rate adiabaticity threshold, one concludes that adiabatic transfer is feasible within the range, 
\begin{equation}
\frac{\xi_s}{t_s}  <  \dot{x} < \frac{1}{3\pi} K~.
\label{apc}
\end{equation}
The upper limit is required for 96\% efficiency \cite{Vitanov17} and ensures small probability for non-adiabatic transitions in the transverse (energy) direction. 

\begin{figure}[t] 
\centering
\includegraphics[width=\hsize]{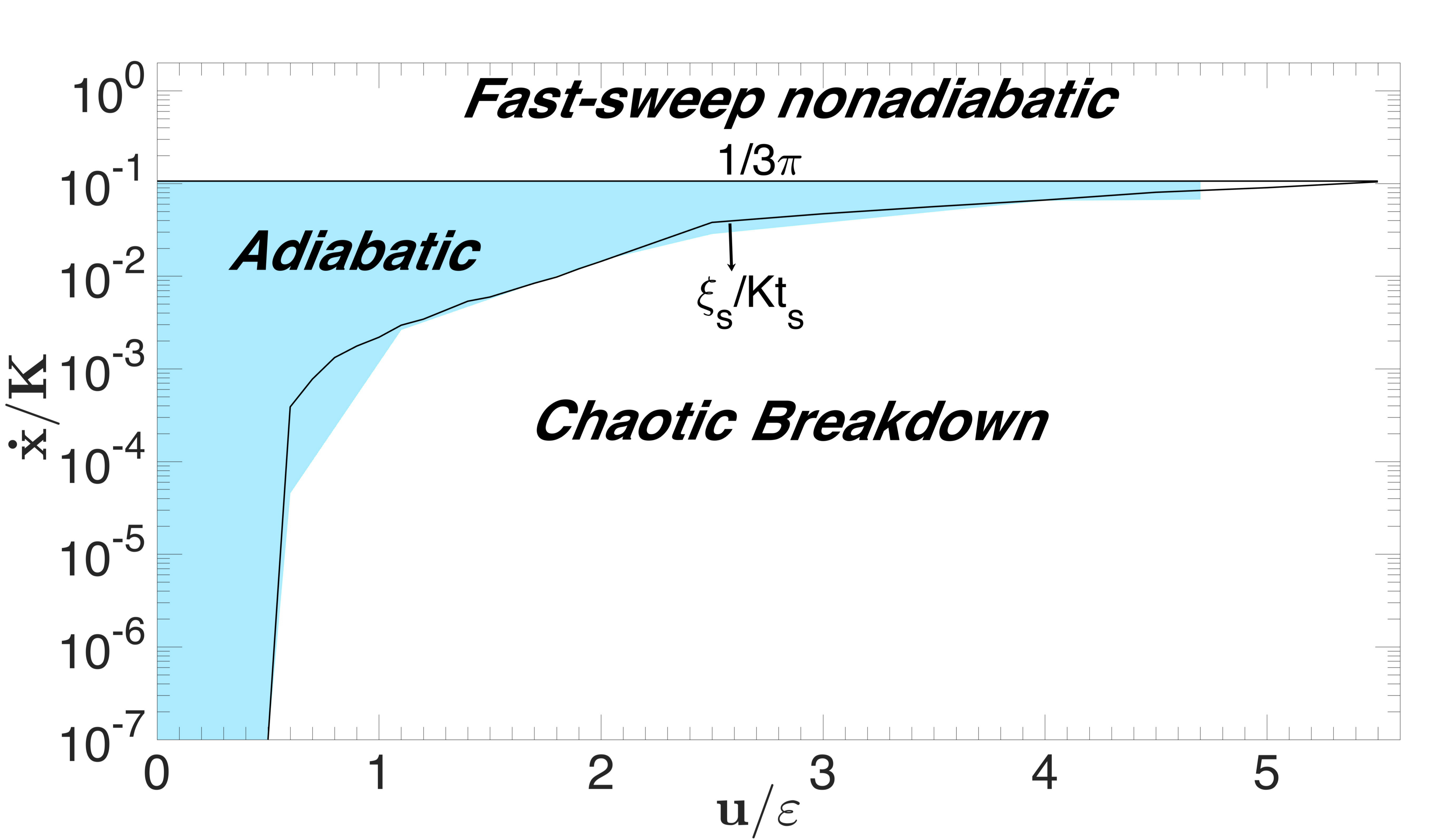} 
\caption{(color online) 
Adiabaticity diagram: The shaded blue region corresponds to the range of parameters 
where the numerically-determined transfer efficiency to the target state $P_3(\infty)$ 
is greater than 96\%. Solid lines mark the predicted boundaries of \Eq{apc}.}
\label{efficiency}
\end{figure}

The dependence of the parameters $\xi_s$ and $t_s$ on $u$ has been studies in~\cite{Dey18}. As the interaction strength increases, the chaotic interval's width grows, while its instability time becomes shorter. Thus, the lower adiabaticity threshold increases  monotonically with~$u$.   When it becomes larger than the $u$-independent upper threshold, adiabatic passage is no longer possible. These predictions are confirmed by numerical classical simulations, as shown in \Fig{efficiency}. The range of effective adiabatic transfer agrees well with the predicted theoretical boundaries. Whereas the upper part of the diagram is just the well-known timescale-separation criterion, the lower chaotic breakdown region has not been previously considered.

\subsection{Quantum detours around chaos}

\begin{figure}[t]
\centering
\includegraphics[width=\hsize]{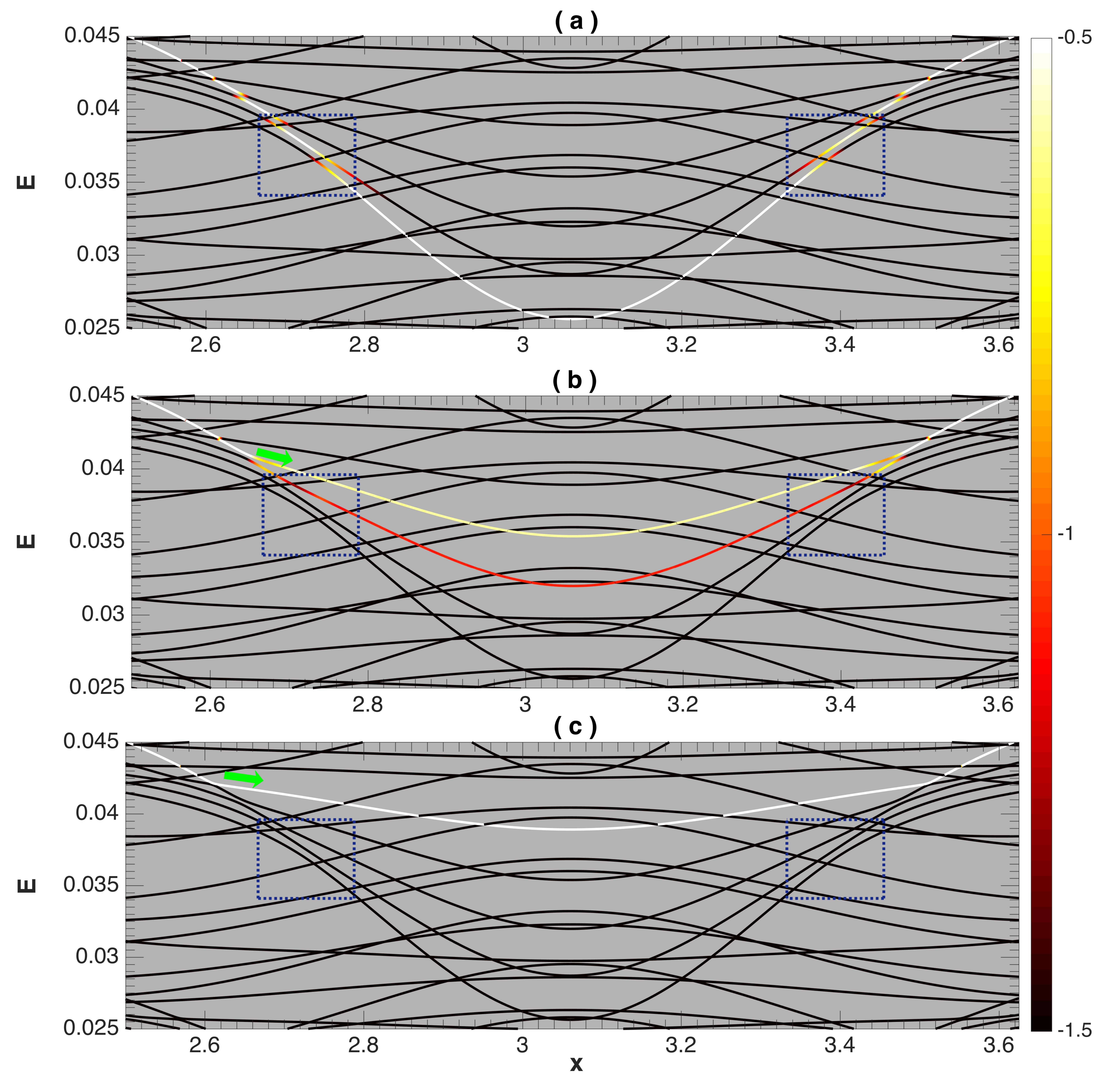} 
\caption{(color online)  
The many-body energy levels $E_{\nu}(x)$ are plotted against~$x(t)$, 
and color-coded according to $p_\nu(t)$.  
Panels~(a) and~(c) are for the simulations 
of \Fig{dynamics}a and \Fig{dynamics}e, respectively, whereas panel (b) is for $\dot x/K=3\times 10^{-4}$.
In panel (a) the system follows the classically adiabatic path, 
formed by a sequence of {\em diabatic} many-body crossings.
In panels (b) and (c), the arrow-marked transitions become adiabatic, 
leading to two different quantum detours.  
The rectangles indicate the levels that are supported 
by the chaotic strip in phase space: their vertical 
edges mark the chaotic interval as in~\Fig{dynamics}. \\ }
\label{pathways}
\end{figure}
%

\begin{figure} [t]
\centering
\includegraphics[width=\hsize]{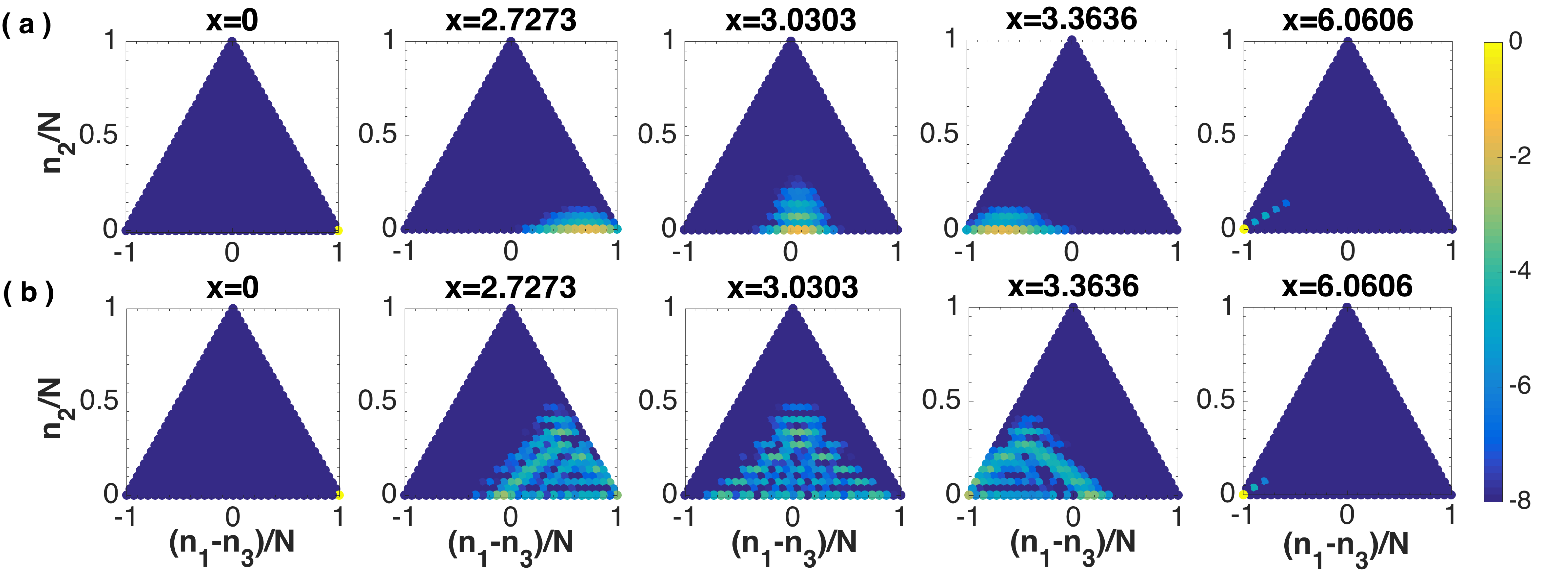} 
\caption{(color online) 
The many-body number distribution $p_{n,n_2}(t)$ 
of the followed states in panels (a) and (c) of  \Fig{pathways}. 
In the classically adiabatic case~(a) 
the many-body state remains coherent, 
while in the quantum detour case~(b) 
it transiently becomes highly non-classical.}
\label{pn}
\end{figure}

While the appearance of low sweep rate boundaries is explained by the semiclassical analysis of Section~III.B, the recovery of STIRAP efficiency at even lower sweep rates in the many-body results of Section III.A, is a pure quantum effect. Its mechanism becomes clear by inspection of the $p_{\nu}$ distributions in \Fig{pathways}, which are essentially a zoom-in on the middle panels of \Fig{dynamics} (Note the vertical axis in \Fig{pathways} is the energy rather than the level-index as in \Fig{dynamics}).  
The STIRAP process  in \Fig{pathways}a (zoom-in of \Fig{dynamics}a) follows the classical SP, i.e. the many body-state traces the classical path in  \Fig{coherence_rep}a, via a series of {\em diabatic} hops between the many body states. This correspondence between the classical-adiabatic and the quantum-diabatic  paths  is also true for the many-body Landau-Zener crossing in the Bose-Hubbard two mode system \cite{Mannschott09}. 
In comparison, the quantum STIRAP recovery shown in \Fig{pathways}b and \Fig{pathways}c (zoom-in of \Fig{dynamics}e), is obtained when the sweep becomes slow enough to make 
one of these crossings (marked arrows) adiabatic. 
The evolution then proceeds diabatically through a series of {\em non-classical} states until the symmetrically twin transition is encountered on the exit, 
and the system returns to the classical path. 

The many-body dynamics thus offers a {\em quantum detour} that avoids the chaotic potholes. This detour has no classical equivalent. Its non-classicality is evident when we compare the Fock number distribution $p_{n,n_2}=|\langle n, n_2|\Psi\rangle|^2$ obtained at different stages of the classical adiabatic path (\Fig{pn}a) and of the quantum detour (\Fig{pn}b). The classical distribution corresponds to a coherent state moving from the initial mode to the target mode. By contrast, the number distribution of the quantum detour state is highly nonclassical and projects substantially onto the intermediate state.

\begin{figure} [t]
\centering
\includegraphics[width=\hsize]{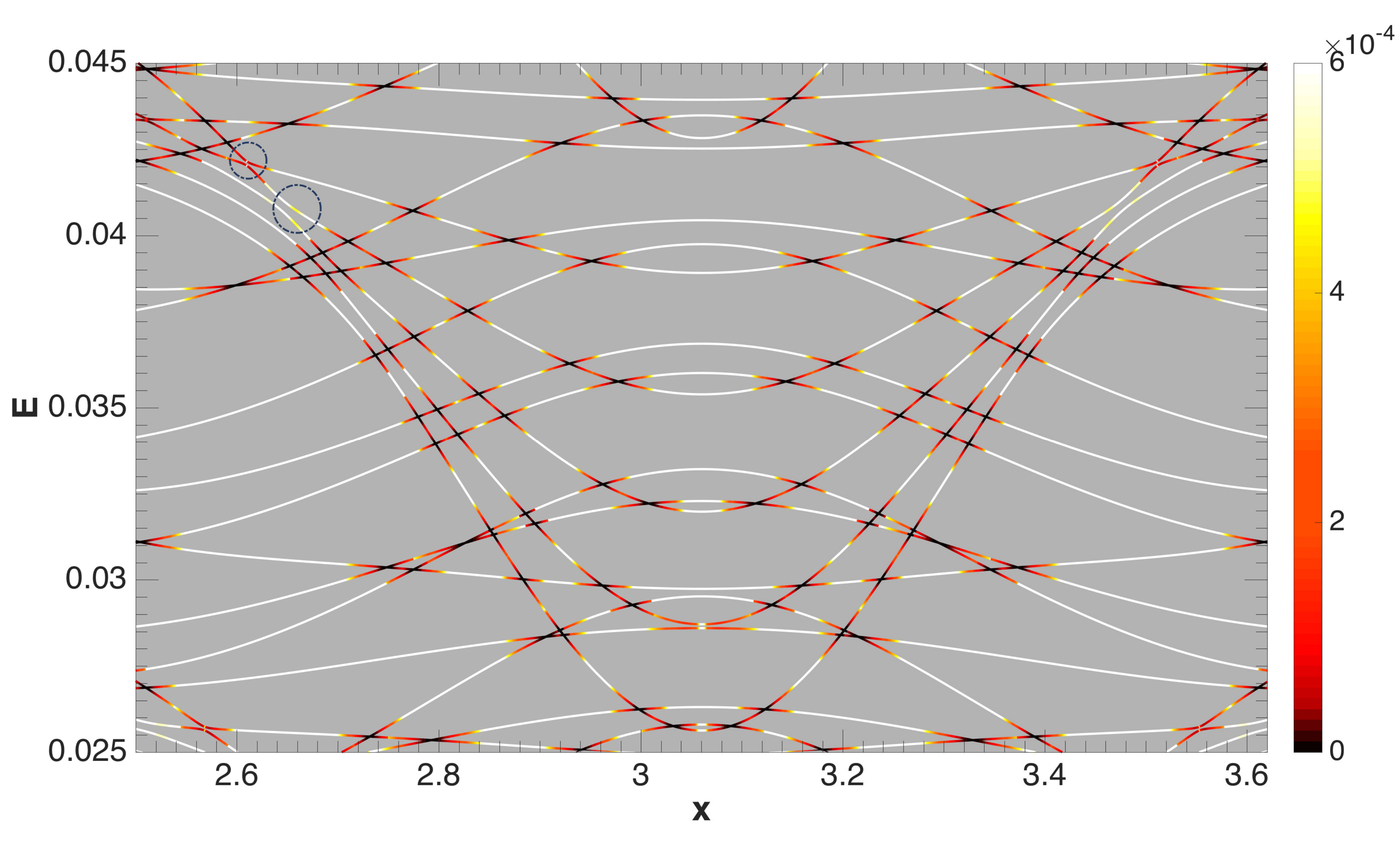} 
\caption{(color online) 
The adiabaticity threshold $\Delta_{\nu,\nu'}^2/\sigma_{\nu,\nu'}$, with $\sigma_{\nu,\nu'}$ approximated as $\partial\Delta_{\nu,\nu'}/\partial x$, for the avoided crossings between the many-body adiabatic eigenstates. The two crossings with the highest threshold, marked by circles, are the ones for which a diabatic~$\rightarrow$~adiabatic transition opens up the quantum detours in our numerical simulations.}
\label{thresh}
\end{figure}

Returning to \Fig{summary}, the high-efficiency adiabatic region is classically limited by the boundaries of \Eq{apc}, as discussed in Section III.B. The corresponding quantum process exhibits another high efficiency region below the sweep rate for which the critical many-body crossing becomes adiabatic and the detour opens up. The threshold for opening a quantum detour is thus obtained from the standard linear curve crossing prescription:
\begin{equation}
%
{\dot x} \ \ < \ \ \frac{\Delta_{\nu,\nu'}^2}{\sigma_{\nu,\nu'}}, \ \ \ \ \ 
\sigma_{\nu,\nu'} \equiv \left| \BraKet{\nu} {\frac{\partial \mathcal{H}}{\partial x}} {\nu'} \right| 
\label{mba}
\end{equation}
where $|\nu\rangle$ and $|\nu'\rangle$ are the participating states and $\Delta_{\nu,\nu'}$ is the energy gap at the avoided crossing.
This should be contrasted with the one-body (``classical") upper limit of \Eq{apc}, where the relevant scales are ${\Delta \sim \sigma \sim K}$.  In \Fig{thresh} we plot the adiabaticity threshold of \Eq{mba} for the various avoided crossings, approximating $\sigma_{\nu,\nu'}$ from their slopes. Indeed, the two crossings with the highest threshold are the ones for which quantum detours are opened in \Fig{pathways}b and  \Fig{pathways}c. In between the thresholds  the PN exhibits an erratic dependence on the sweep rate, which can be regarded as the time domain version of universal conductance  fluctuations. It is due to the interference between the various detour pathways that are available for the evolution.

\section{conclusions}

The classical dynamics of nonlinear STIRAP is strongly affected by dynamical chaos. For repulsive interaction, it is chaos-induced dynamical instability, rather than energetic instability,  that causes the nonlinear failure of adiabatic passage  and sets novel low sweep rate boundaries for efficient adiabatic transfer. 
Going beyond the classical picture, the immense increase in state-space dimensionality from the 4D classical phase-space to the $\sim N^2/2$ dimensional quantum Hilbert space, opens new quantum avenues for adiabatic passage that do not exist in the restricted classical picture.
Thus, quantum dynamics offers alternative many-body pathways that circumvent chaos and reenable adiabatic passage in regions where it is classically forbidden. These quantum detours have no classical equivalent and are therefore not based on the classical dark state. They should thus be considered as separate many-body adiabatic passage schemes rather than a modification of the classical scenario.

\acknowledgements
This research was supported by the Israel Science Foundation (Grant  No. 283/18)

\appendix

\section{The semiclassical Hamiltonian}

The mean-field limit of the interacting many-body system is attained as $N\rightarrow\infty$ while $UN$ is held fixed.  In this limit, the many-body dynamics may be restricted to the $\gamma=1$ classical coherent states, so that field operators can be replaced by their expectation values $\hat{a}_j\rightarrow\langle \hat{a}_j \rangle\equiv a_j=\sqrt{n_j} e^{i \phi_j}$. The classical Hamiltonian therefore takes the form,
\begin{eqnarray}
{\cal H}_{{\rm{cl}}}&=&{\cal E} n_2-\left[\Omega_p(x) \sqrt{n_1n_2}{\rm cos }(\phi_2 -\phi_1)\right.\nonumber \\
~&~&\left.-\Omega_s(x) \sqrt{n_2n_3}{\rm cos }(\phi_2 -\phi_3)\right]+\frac{U}{2}\sum_{j=1}^3 n^2_j.
\end{eqnarray}
Rescaling the classical amplitudes as $a_j \rightarrow a_j/\sqrt{N}$, and time as $t\rightarrow Kt$, and defining $P_j=|a_j|^2$, we obtain the discrete nonlinear Schr\"odinger equations :
\begin{equation}
i \dot{{\bf a}}=({\mathcal H}_0+u{\mathcal P}){\bf a}~,
\label{classical_eqn}
\end{equation}
where,
\begin{equation}
\mathcal{H}_0=
\begin{pmatrix}
0&-\kappa_p/2&0\\
-\kappa_p/2&\varepsilon&-\kappa_s/2\\
0&-\kappa_s/2&0
\end{pmatrix},~
\mathcal{P}{=}
\begin{pmatrix}
P_1&0&0\\
0&P_2&0\\
0&0&P_3
\end{pmatrix} \ \ \ \ 
\end{equation}

The classical adiabatic basis consists of the SPs of the grand canonical Hamiltonian ${\cal H}_{\rm{cl}}-\mu N$, 
i.e. the classical points that satisfy,
\begin{equation}
i\dot{{\bf a} }= \mu {\bf a}
\end{equation}
where $\mu$ is the chemical potential. The classical adiabatic energies $E(SP)$ are the values of ${\cal H}_{\rm{cl}}$ at these classical SPs. For $u=0$, they are just the three well-known eigenenergies of the linear STIRAP scenario \cite{Gaubatz90, Vitanov17}. The followed SP is the dark state, 
\beq
\bm{a}_{\mathrm{sp}}(x)=(\cos\vartheta(x),0,-\sin\vartheta(x))~,
\eeq
whose energy is  $E(SP)=0$. With non-zero interaction, the SPs are shifted and bifurcate if the effective interaction $u_{\text{eff}}(x)$ is sufficiently strong, as seen in \Fig{spectra}b. One such bifurcation is the 'horn' avoided crossing  \cite{Graefe06} which appears for $u>\varepsilon$. More SPs, up to a max total of eight, emerge as $u$ is increased.

\begin{figure}[t]
   \centering
   \includegraphics[width=\hsize]{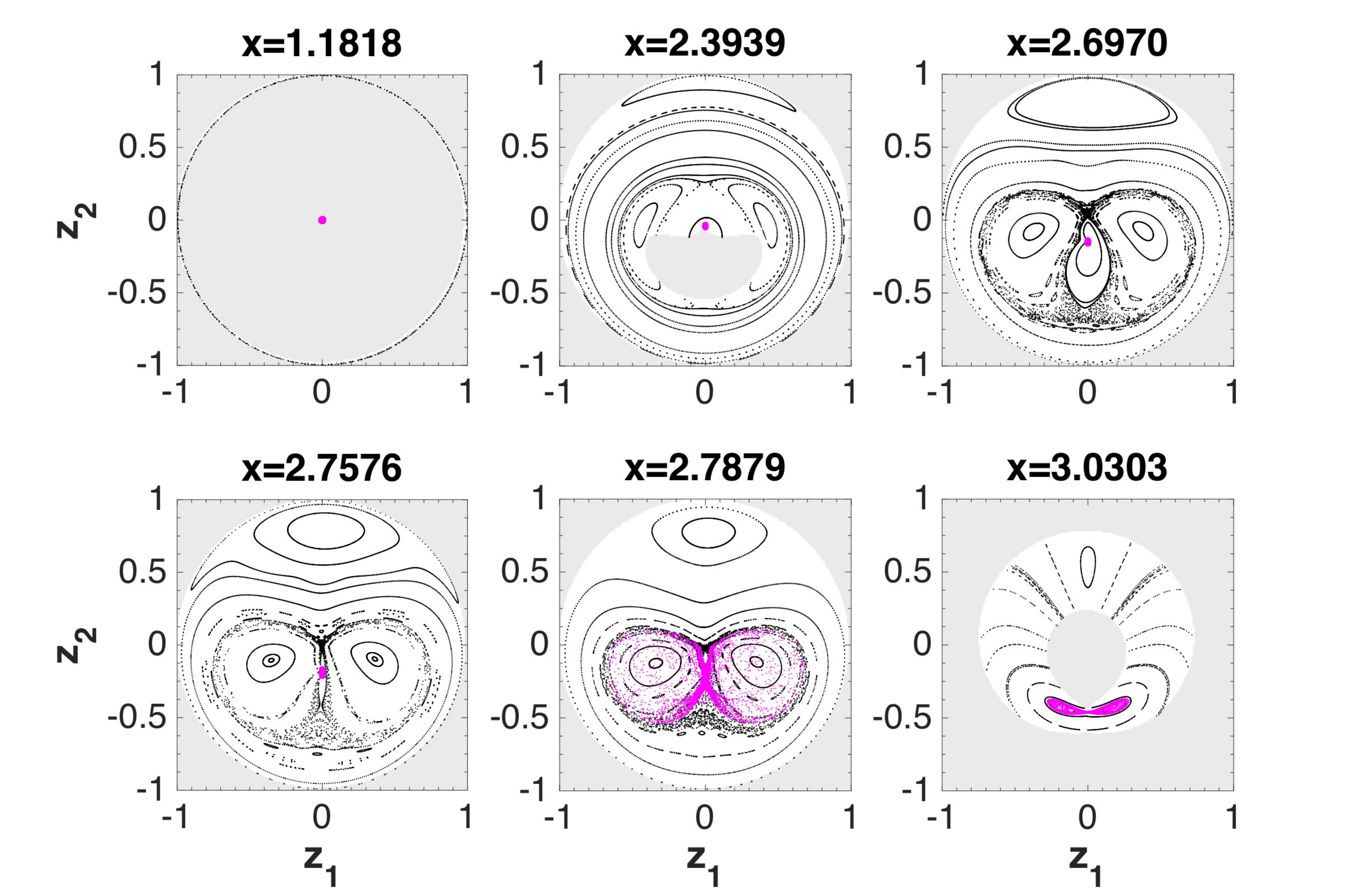} 
   \caption{(color online) Poincare sections for the frozen Hamiltonian at representative values of $x$ for $u=2.2\varepsilon=0.22$. The energy in all panels is $E=E(SP)$ of the followed SP. The cross-sections are taken through the $n_2=n_2[SP]$ plane of the 3D energy surface. We use polar coordinates $z_1=r\sin\varphi$, $z_2=r\cos\varphi$ where $r=[1-(n/N)]/2$. Magenta dots correspond to a semiclassical cloud, initially localized around the followed SP. Gray shading marks energetically forbidden regions. Note that these panels depict the adiabatic sequence up to the middle point $x\sim3$. The Poincare sections at later times mirror the presented panels. \\ }
   \label{pcp}
\end{figure}
%
%

\section{Bogoliubov stability analysis}

In order to determine the stability of the followed classical SP ${\bf a}_{\mathrm{sp}}$ at any $x$, we employ the Bogoliubov formalism.  Linearizing ${\bf a}={\bf a}_{\mathrm{sp}}+\delta{\bf a}$ in \Eq{classical_eqn}, retaining only linear terms in $\delta{\bf a}$, and carrying out the Bogoliubov transformation,
\begin{equation}
{\bf \delta a}=e^{-i\mu t}\left({\bf u} e^{-i\omega t} - {\bf v}^* e^{i\omega t} \right)~,
\end{equation}
we obtain the discrete Bogoliubov equations,
\begin{eqnarray}
\left(
{\mathcal H}_0+2u{\mathcal P}_{\mathrm{sp}}
-\mu-\omega\right) {\bf u}
-u{\mathcal P}_{\mathrm{sp}}{\bf v}&=&0~,\nonumber\\
\left(
{\mathcal H}_0+2u{\mathcal P}_{\mathrm{sp}}
-\mu+\omega\right) {\bf v}
-u{\mathcal P}_{\mathrm{sp}}{\bf u}&=&0~,
\label{bogeq}
\end{eqnarray}
where ${\mathcal P}_{\mathrm{sp}}$ is the occupation matrix ${\mathcal P}$ at ${\bf a}={\bf a}_{\mathrm{sp}}$. For a system with $d$ degrees of freedom, there are $d$ collective modes $\{{\bf u}_j, {\bf v}_j\}_{j=1,..,d}$ with corresponding characteristic frequencies $\omega_j$. Our trimer model has in principle $d=3$ (three $n_j,\varphi_j$ pairs serving as conjugate action-angle coordinates). However, due to the conservation of the total particle number $N$, the system is invariant under global phase transformations, leaving only two degrees of freedom (two population imbalances and two relative phases serving as action angle variables). Accordingly, the Bogoliubov frequencies include  a zero mode $\omega_0=0$ and two non-vanishing frequencies $\omega_{1,2}$. 
Defining the quasiparticle operators,
\begin{equation}
\hat{c}_j={\bf u}_j\cdot\hat{\bf a} + {\bf v}_j\cdot\hat{\bf a}^\dag~,
\end{equation}
transforms the many-body Hamiltonian of \Eq{ham}  in the vicinity of the SP, into the approximate quadratic form in \Eq{hamq}.

\begin{figure}[t]
   \centering
   \includegraphics[width=\hsize]{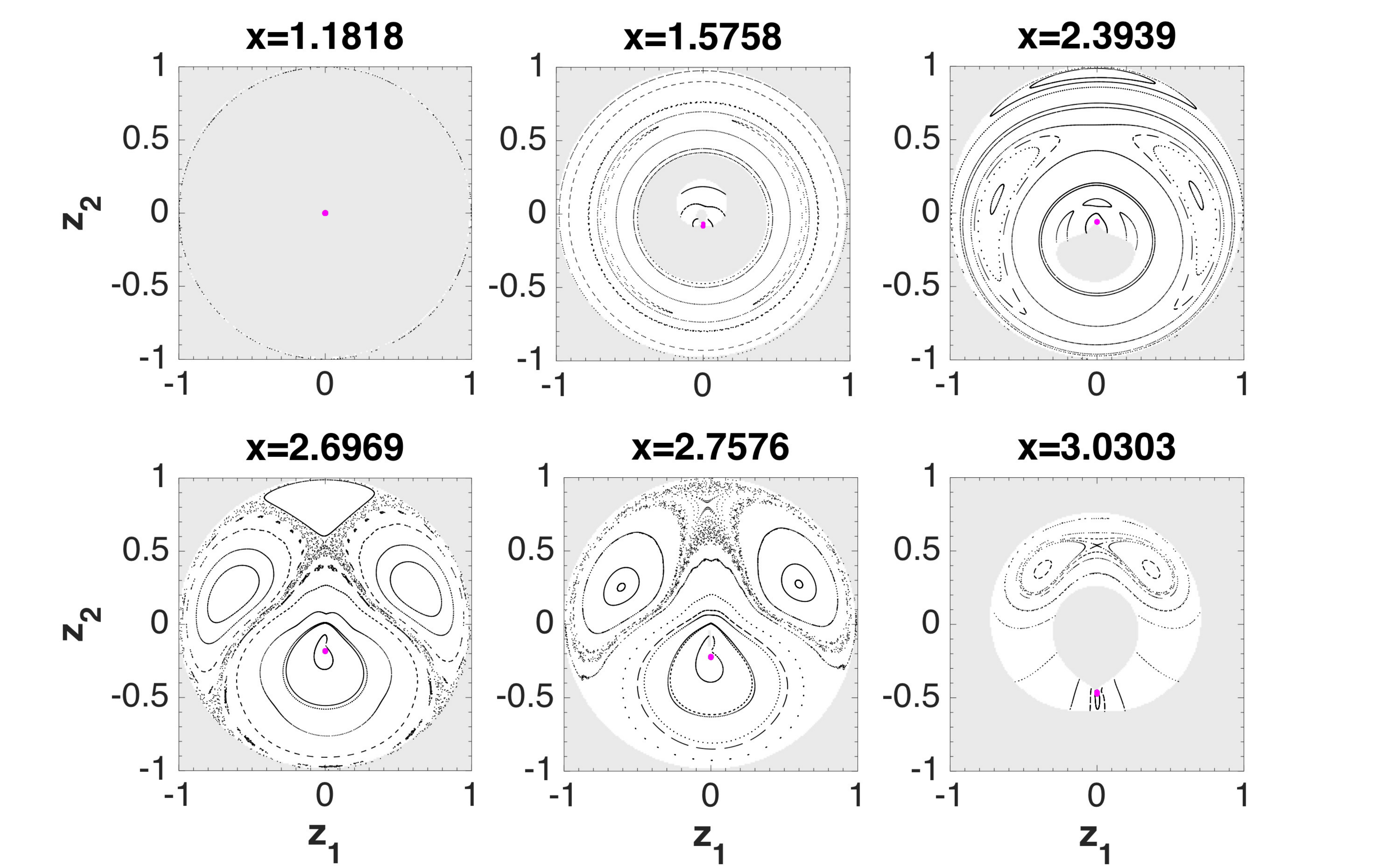} 
   \caption{(color online) Same as \Fig{pcp} for attractive interaction: $u=-2.2\varepsilon=-0.22$.}
   \label{pcn}
\end{figure}

\section{Poincare sections}

The source of dynamical instability is readily seen by plotting quasistatic (frozen-$x$) Poincare sections during the STIRAP sequence. The fixed energy surfaces within the 4D phasespace of the Bose-Hubbard trimer are three dimensional. For a given $N$ and $E$ our dynamical coordinates are accordingly $n_2$, $n$, and the relative phase ${\varphi=\varphi_1{-}\varphi_3}$. The Poincare section consists of trajectories in the energy surface of the followed SP, $E=E[\text{SP}]$. A trajectory is sampled each time that it intersects the plane ${n_2=n_2[\text{SP}]}$. We thus obtain a section whose coordinates are ${\mathbf{z}=(\varphi,n)}$. These are displayed as polar coordinates in \Fig{pcp} and \Fig{pcn}. The Poincare sections map is thus a Lambert conical projection of a sphere, where the radii are the meridians and the azimuth is the longitude. The origin corresponds to the initial state ($P_1=1$) that lies on the north pole  and the $z=1$ circumference corresponds to the target state ($P_3=1$) that lies on the south pole. 
Note that the observed structures do not reflect the topography of the energy landscape, but correspond to various periodic orbits, invariant tori, and chaotic regions {\em on the same energy surface}. The plotted sections contain a single SP that supports the followed adiabatic eigenstate, while the other 'fixed-points' are in fact periodic orbits. In each section, we plot the distribution obtained by classically evolving a cloud of classical trajectories, initially localized  around the followed-SP.

\begin{figure}[t]
\centering
\includegraphics[width=\hsize]{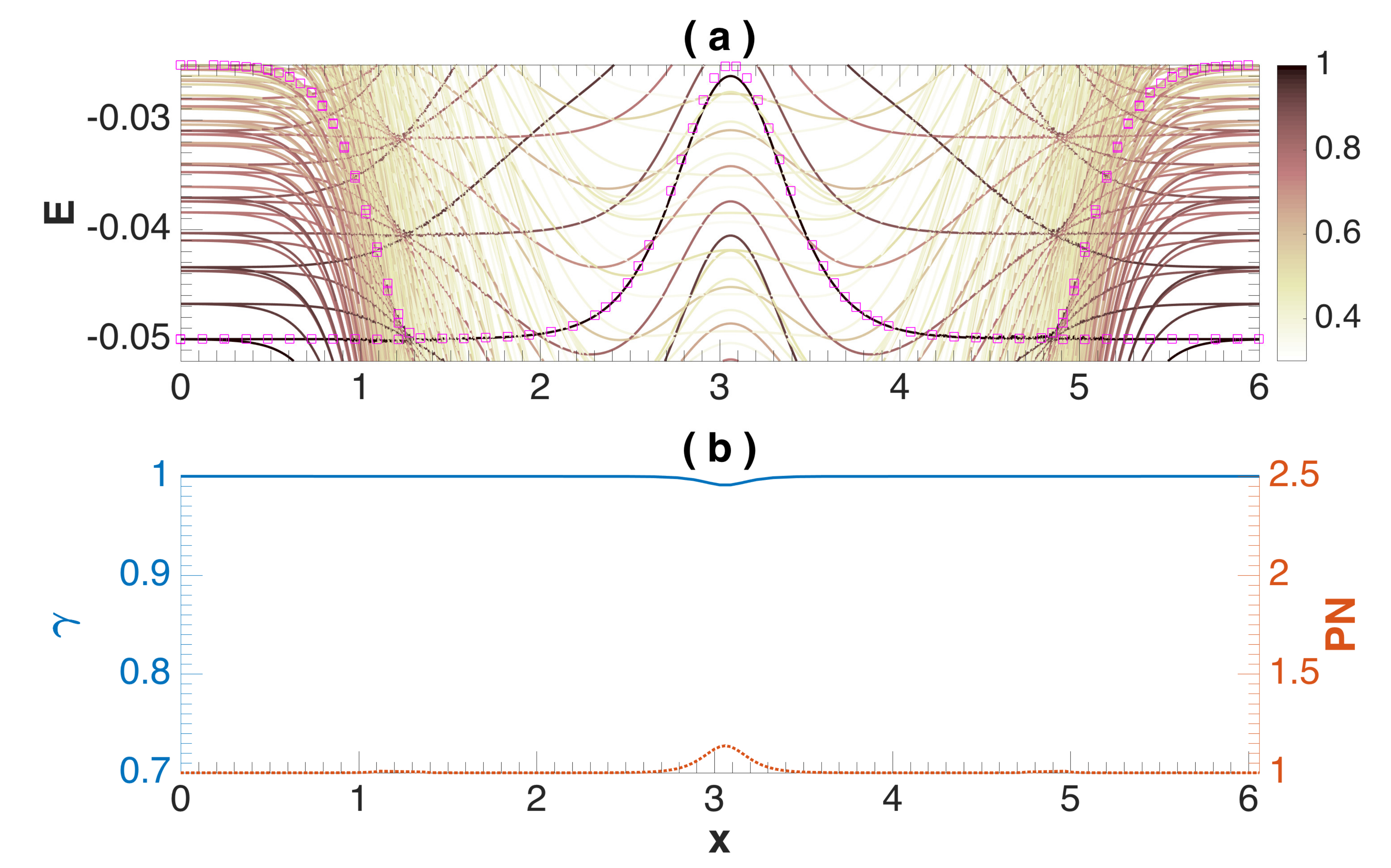} 
\caption{(color online) Same as \Fig{coherence_rep} for attractive interaction $u=-0.1$.}
\label{coherence_att}
\end{figure}

The sequence of Poincare sections in \Fig{pcp} represents the case of repulsive interaction. At early times ($x=1.1818$), the followed SP is an energy maximum near $(n,n_2)=(N,0)$ (the origin), surrounded by a lower energy forbidden region (gray). It is degenerate with the other self-trapped maxima near $(n,n_2)=(-N,0)$ (the circumference)  and $(n,n_2)=(0,N)$ (lying outside the ${n_2=n_2[\text{SP}]}$ section). The followed SP becomes an energy saddle after the horn crossing. Hence the forbidden region disappears, and an intermediate non-linear resonance appears as a `belt' in the Poincare section ($x=2.3939$). 
At larger~$x$, the belt expands, and a chaotic strip is formed along its border  ($x=2.6970$). The enclosed 'island', containing the followed-SP, shrinks down until the SP enters the chaotic strip ($x=2.7576$). The dynamical instability thus corresponds to the embedding of the followed-SP in the chaotic strip, resulting in the quasi-stochastic spreading of the initially localized distribution over the chaotic region ($x=2.7879$). 
The entire progression takes place on a single 3D energy surface, and has no trace in the adiabatic energy diagram.

\begin{figure}[b]
   \centering
   \includegraphics[width=\hsize]{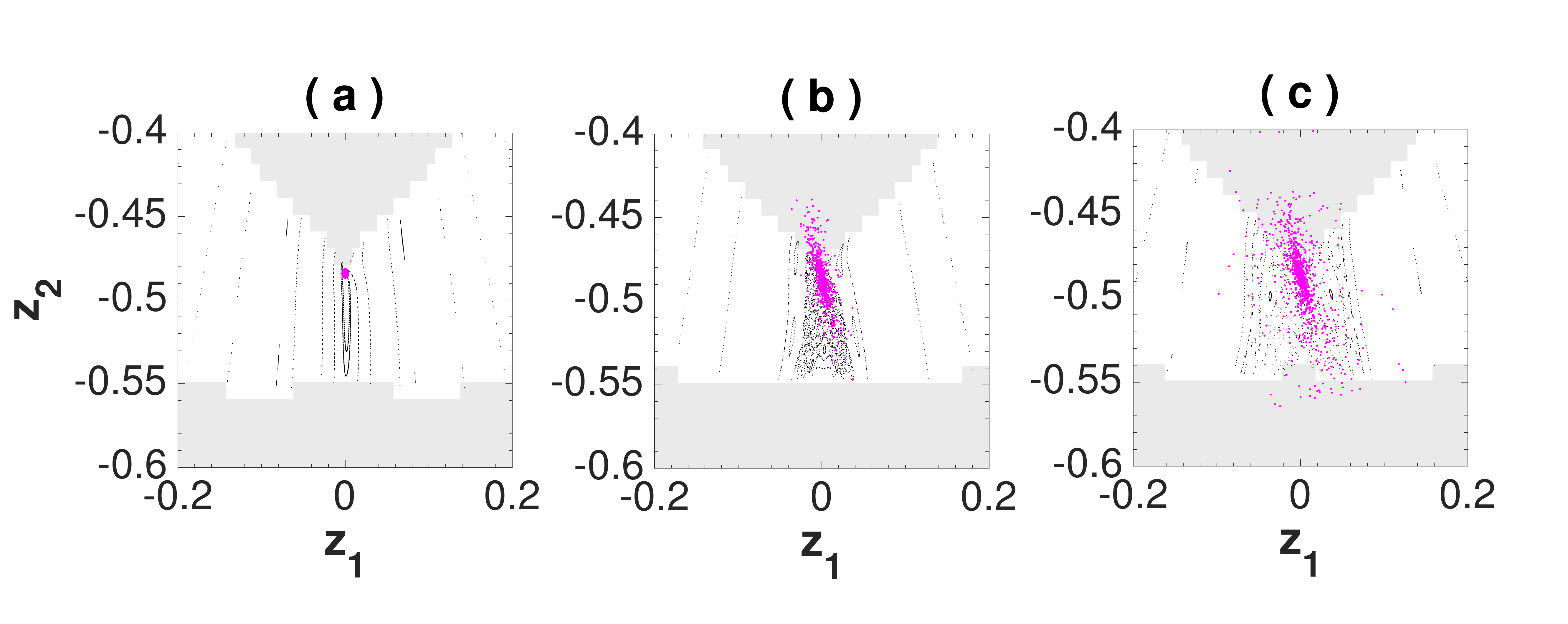} 
   \caption{(color online) Close up on the followed SP region of the Poincare sections at $x=3.0454$ for: (a) $u=-2.2\varepsilon$; (b) $-3.5\varepsilon$; (c) $-3.8\varepsilon$.}
   \label{pcnb}
\end{figure}

The situation is quite different for attractive interaction, see \Fig{pcn}. The early time SP is a minimum since the dominant attractive interaction favors the localization of all particles in one of the three modes. The transition  to a saddle-point takes place at the horn crossing (present at any $u<0$) and a chaotic belt appears here too. However, for $u>-2.2\varepsilon$ the followed SP is located away from the chaotic region at all time and therefore never loses its dynamical stability. Consequently, chaotic intervals do not exist for nonlinear STIRAP with attractive interaction in this range, and the only obstacle to successful adiabatic passage is the traverse of the horn crossing. From the quantum perspective, the absence of a chaotic interval is manifested in the one-particle coherence maintained by the many-body eigenstates along the entire classical path, see \Fig{coherence_att} in comparison to \Fig{coherence_rep}. \rmrk{The dynamical instability does reappear for sufficiently strong ($u<-2.2\varepsilon$) attractive interaction (see \Fig{bog_r}f, the dash-dotted line in \Fig{bog_i}, and the Poincare sections in \Fig{pcnb}).}

\begin{figure}[t]
\centering
\includegraphics[width=\hsize]{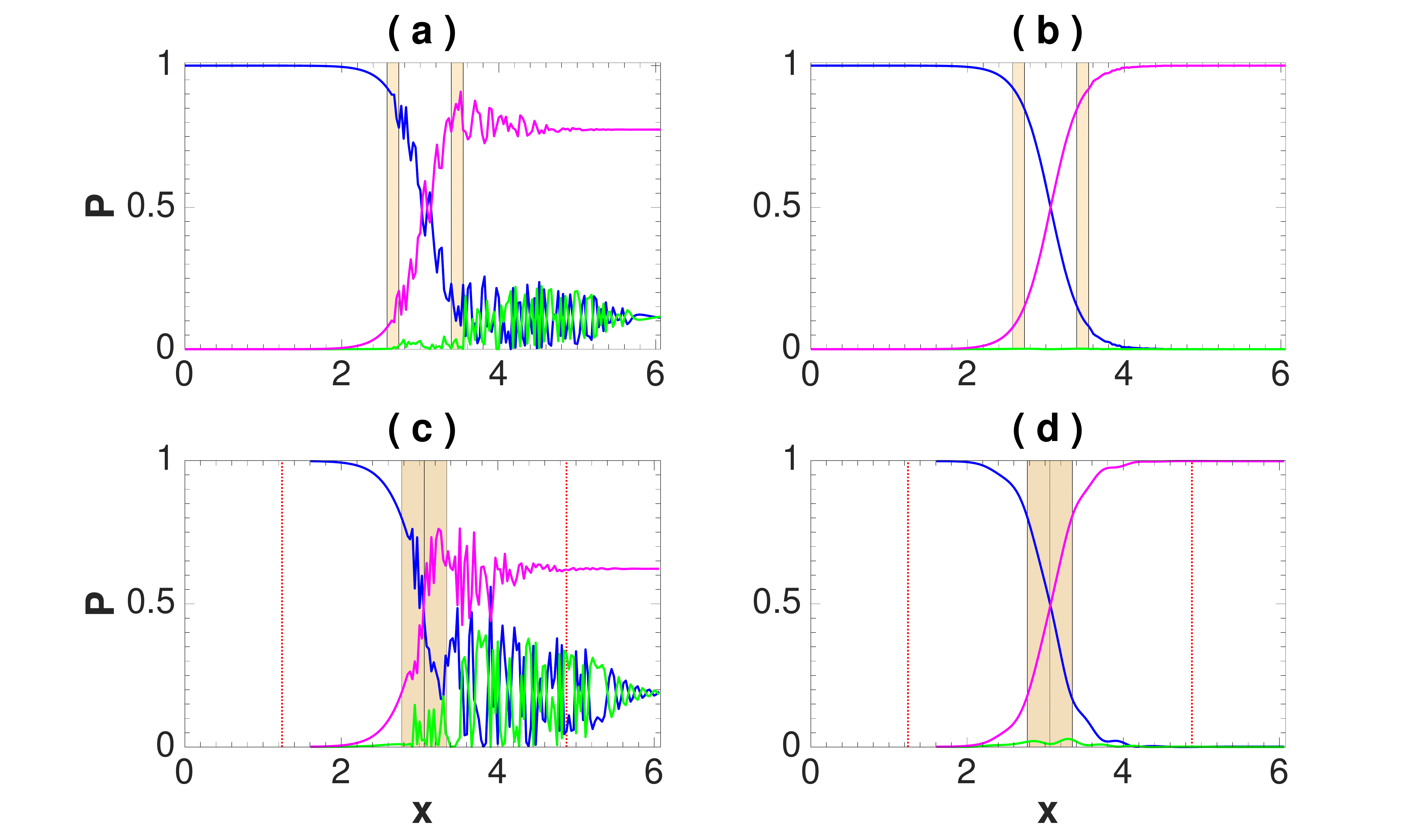} 
\caption{(color online) Evolution of the site populations for repulsive interaction. {\bf (a)}~Failure of STIRAP in the absence of SP bifurcations: here $u=0.8\varepsilon$ is {\em below} the critical value for obtaining the horn crossing. The sweep rate is $\dot{x}/K=6\times10^{-5}$.  {\bf (b)}~Recovery of adiabatic passage with {\em increased} sweep rate ($\dot{x}/K=6\times10^{-4}$) during chaotic intervals. {\bf (c)}~~Failure of STIRAP for $u=2\varepsilon$, with initial conditions that bypasses the horn crossing: the process is launched at the adiabatic state {\em after} the avoided crossing. 
Sweep rate is $\dot{x}/K=6\times10^{-4}$. {\bf (d)}~For same $u$, efficiency is recovered due to faster sweep ($\dot{x}/K=4\times10^{-2}$). \\ }
\label{pdp}
\end{figure}
%
%

\section{Semiclassical STIRAP efficiency}

For repulsive interaction ($u>0$), the cause for slow-sweep breakdown of adiabaticity in semiclassical simulations, is the dynamical instability (i.e. chaos) rather than energetic instability \cite{Dey18}. For example, In \Fig{pdp}a, adiabaticity breaks down even if no avoided crossing is present and the followed SP remains an energy maximum throughout its evolution.  Nonadiabatic population oscillations are clearly boosted during the marked chaotic intervals.  Moreover, as shown in Fig~\Fig{pdp}c, while for ${u>\varepsilon}$ the horn crossing does appear in an early stage, adiabaticity breaks down even if it is bypassed by initiating the system {\em after} it. 

\begin{figure}[t]
\centering
\includegraphics[width=\hsize]{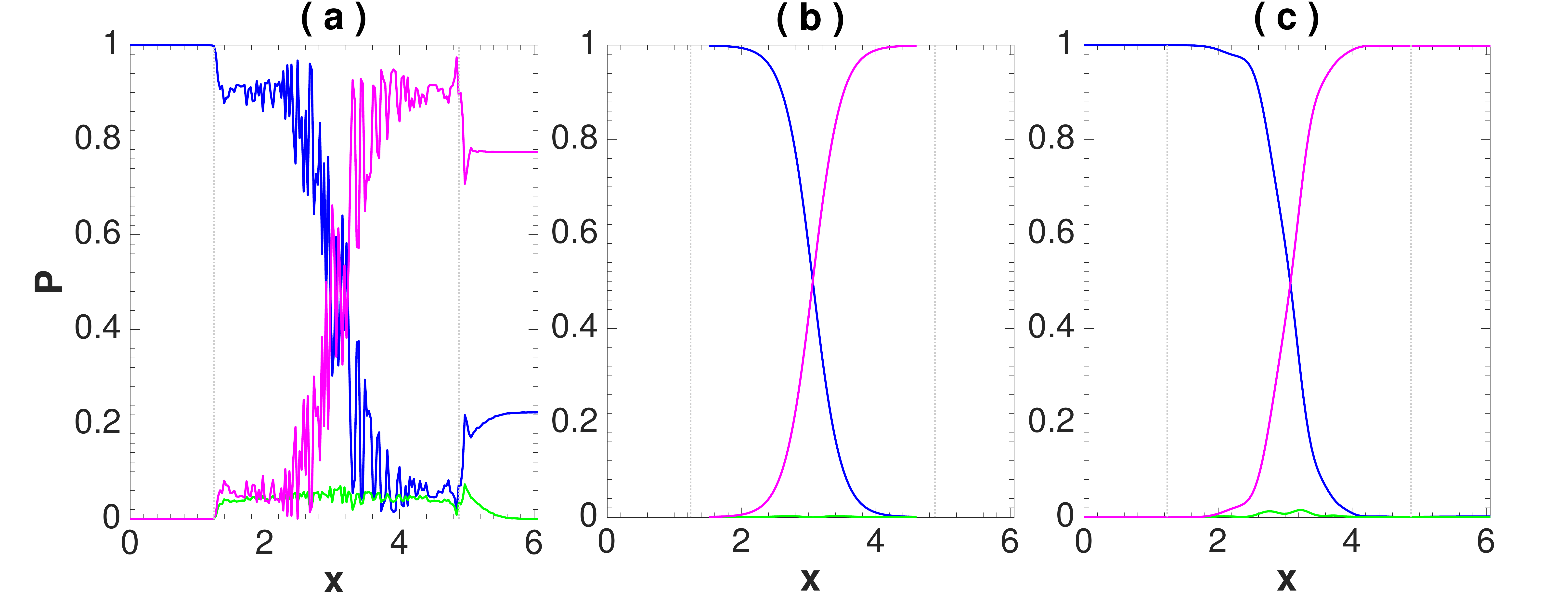} 
\caption{(color online) {\bf (a)} Site populations dynamics for attractive interaction, $u=-\varepsilon=-0.1$ and $\dot{x}/K=6\times10^{-6}$.  {\bf (b)} Same, launching the system after the first horn crossing and stopping before the second. {\bf (c)} Same, for $\dot{x}/K=4\times10^{-2}$}
\label{pdn}
\end{figure}

The situation is entirely different for attarctive interaction ($u<0$) as illustrated in \Fig{pdn}. Here, the nonlinear breakdown of adiabaticity takes place immediately at the horn crossing rather than later (see \Fig{pdn}a). Since no chaotic intervals are encountered by the followed SP, launching the system after the horn crossing recovers adiabaticity (see \Fig{pdn}b), in marked contrast to the repulsive interaction scenario (see \Fig{pdp}c). The cause of failure when the interaction is attractive is thus the inability of the system to {\em diabatically} traverse the avoided crossing from the pre-horn minimum to the post-horn saddle. Here too, the remedy is a faster sweep, allowing for such diabatic crossing  (see \Fig{pdn}c). 

In principle, a horn-crossing effect exists also for repulsive interaction. However, to see failure due to the horn crossing with $u>0$,  the sweep rate needs to be well below the threshold  for chaotic failure, specified in Eq.(~\ref{apc}). Thus, for repulsive interaction, any horn effect is overwhelmed by the passage through chaos mechanism. The absence of the latter for attractive interaction, allows for the observation of the horn-crossing breakdown.

\begin{thebibliography}{30}
\bibitem{Landau32}
L. D. Landau, 
Phys. Z. Sowjetunion {\bf 2}, 46 (1932).

\bibitem{Zener32}
C. Zener,
Proc. R. Soc. London, ser. A  {\bf137},  696 (1932).

\bibitem{Stuckelberg32}
E. C. G. Stuckelberg,
Helv. Phys. Acta {\bf 5}, 369 (1932).

\bibitem{Majorana32}
E. Majorana,
Nuovo Cimento {\bf 9}, 45 (1932).

\bibitem{Gaubatz90}
Gaubatz, U., P. Rudecki, S. Schiemann, and K. Bergmann,
J. Chem. Phys. {\bf 92}, 5363 (1990).

\bibitem{Vitanov17}
N. V. Vitanov, A. A. Rangelov, B. W. Shore, and K. Bergmann,
Rev. Mod. Phys. {\bf 89}, 015006 (2017).

\bibitem{Zobay00}
B. M. Zobay and B. M. Garraway, Phys. Rev. A {\bf 61}, 033603 (2000).

\bibitem{Wu00}
B. Wu and Q. Niu, 
Phys. Rev. A {\bf 61}, 023402 (2000).

\bibitem{Liu02}
J. Liu, L.-B. Fu, B.-Yi. Ou, S.-G. Chen, D.-I. Choi, B. Wu, and Q. Niu
Phys. Rev. A 66, 023404 (2002)

\bibitem{Liu03}
J. Liu, B. Wu and Q. Niu, 
Phys. Rev. Lett. {\bf 90}, 170404 (2003).

\bibitem{Witthaut06}
D. Witthaut, E. M. Graefe, and H. J. Korsch,
Phys. Rev. A  {\bf 73}, 063609 (2006)

\bibitem{Witthaut11}
D. Witthaut, F. Trimborn, V. Kegel, and H. J. Korsch,
Phys. Rev. A 83, 013609 (2011).

\bibitem{Ishkhanyan10}
A. M. Ishkhanyan,
Europhys. Lett.  {\bf 90}, 30007 (2010).

\bibitem{Altland08}
A. Altland and V. Gurarie
Phys. Rev. Lett. {\bf 100}, 063602 (2008).

\bibitem{Trimborn10}
F Trimborn, D Witthaut, V Kegel, and H J Korsch,
New J. Phys. {\bf 12}, 053010 (2010).

\bibitem{Mannschott09}
K. Smith-Mannschott, M. Chuchem, M.Hiller, T. Kottos, and D. Cohen
Phys. Rev. Lett. {\bf 102}, 230401 (2009).

\bibitem{Chen11}
Y.-A. Chen, S. D. Huber, S. Trotzky, I. Bloch, and E. Altman,
Nature Physics {\bf 7}, 61 (2011).

\bibitem{Paredes13}
S. F. Caballero-Bentez and R. Paredes,
Phys. Rev. A 85, 023605 (2013).

\bibitem{Javanainen99}
J. Javanainen and M. Mackie
Phys. Rev. A {\bf 59}, R3186(R) (1999).

\bibitem{Yurovsky00}
V. A. Yurovsky, A. ben-reuven, P. Julienne, and C. J. Williams,
Phys. Rev. A {\bf 62}, 043605 (2000).

\bibitem{Heinzen00}
D. J. Heinzen, Roahn Wynar, P. D. Drummond, and K. V. Kheruntsyan,
Phys. Rev. Lett. {\bf 84}, 5029 (2000).

\bibitem{Ishkanyan04}
A. Ishkhanyan, M. Mackie, A. Carmichael, P. L. Gould, and J. Javanainen,
Phys. Rev. A 69, 043612 (2004).

\bibitem{Altman05}
E. Altman and A. Vishwanath,
Phys. Rev. Lett. 95, 110404 (2005).

\bibitem{Pazy05}
E. Pazy, I. Tikhonenkov, Y. B. Band, M. Fleischhauer, and A. Vardi,
Phys. Rev. Lett. {\bf 95}, 170403 (2005).

\bibitem{Tikhonenkov06}
I. Tikhonenkov, E. Pazy, Y. Band, M. Fleischhauer, and A. Vardi,
Phys. Rev. A 73 043605 (2006).

\bibitem{liu08}
J. Liu, B. Liu, and L. B. Fu,
Phys. Rev. A {\bf 78}, 013618 (2008).

\bibitem{liu08b}
J. Liu, L.-B. Fu, B. Liu, and B. Wu,
New J. Phys. 10 123018 (2008).

\bibitem{Graefe06}
E. M. Graefe, H. J. Korsch, and D. Witthaut,
Phys. Rev. A {\bf 73}, 013617 (2006).

\bibitem{Rab08}
M. Rab, J. H. Cole, N. G. Parker, A. D. Greentree, L. C. L. Hollenberg, and A. M. Martin,
Phys. Rev. A {\bf 77}, 061602(R) (2008).

\bibitem{Bradly12}
C. J. Bradly, M. Rab, A. D. Greentree, and A. M. Martin,
Phys. Rev. A 85, 053609  (2012).

\bibitem{Polo16}
J. Polo, A. Benseny, Th. Busch, V. Ahufinger, and J. Mompart,
New J. Phys. {\bf 18}, 015010 (2016).

\bibitem{Dupont-Nivet15}
M. Dupont-Nivet, M. Casiulis, T. Laudat, C. I. Westbrook, and S. Schwartz,
Phys. Rev. A 91, 053420  (2015).

\bibitem{Dey18}
A. Dey, D. Cohen, and A. Vardi, 
Phys. Rev. Lett. {\bf 121}, 250405 (2018).

\bibitem{Arwas15}
G. Arwas, A. Vardi, and D. Cohen, 
Sci. Rep. {\bf 5}, 13433 (2015).

\bibitem{Eilbeck95}
J. C. Eilbeck, G. P. Tsironis, and S. K. Turitsyn,
Phys. Scr. {\bf 52}, 386 (1995).

\bibitem{Hennig95}
D. Hennig, H. Gabriel, M.F. Jorgensen, P.L. Christiansen, and C.B. Clausen, 
Phys. Rev. E {\bf 51}, 2870 (1995).

\bibitem{Franzosi03}
R. Franzosi, V. Penna,
Phys. Rev. E {\bf 67}, 046227 (2003).

\bibitem{Flach97}
S. Flach and V. Fleurov, 
J. Phys.: Condens. Matter {\bf 9}, 7039 (1997).

\bibitem{Nemoto00}
K. Nemoto, C.A. Holmes, G.J. Milburn, and W.J. Munro,  
Phys. Rev. A {\bf 63}, 013604 (2000).

\bibitem{Franzosi02}
R. Franzosi and V. Penna, 
Phys. Rev. A {\bf 65}, 013601 (2002).

\bibitem{Hiller06}
M. Hiller, T. Kottos, and T. Geisel, 
Phys. Rev. A {\bf 73}, 061604(R) (2006).

\bibitem{Tikhonenkov13}
I. Tikhonenkov, A. Vardi, J. R. Anglin, and D. Cohen,
Phys. Rev. Lett. {\bf 110}, 050401 (2013).


\clearpage
\end{thebibliography}
\end{document}